\documentclass[prb,amsmath,amssymb]{revtex4}

\begin{document}

\title{Phase Integral Approximation for coupled ODEs of the Schr\"odinger type}

\author{Andrzej A. Skorupski}
\email[]{askor@fuw.edu.pl}
\affiliation{Department of Theoretical Physics, Soltan Institute for Nuclear
Studies, Ho\.za 69, 00--681 Warsaw, Poland}


\begin{abstract}
Four generalizations of the Phase Integral Approximation (PIA) to  sets of
ordinary differential equations  of Schr\"odinger type $\bigl(u_j''(x) +
\sum_{k=1}^N R_{jk}(x) \, u_k(x) = 0 \, ,$ $j = 1, 2, \dots , N \, \bigr)$
are described. The recurrence relations for higher order corrections are given
in a form valid to arbitrary order and for the matrix $\mathbf{R}(x) (\equiv
\{ R_{jk}(x) \})$ either hermitian or non-hermitian. For hermitian and
negative definite $\mathbf{R}(x)$ matrices, a Wronskian conserving PIA theory
is formulated which generalizes Fulling's current conserving theory pertinent to
positive definite $\mathbf{R}(x)$ matrices. The idea of a modification of the
PIA, well known for one equation $\bigl( u''(x) + R(x) \, u(x) = 0 \bigr)$ is
generalized to sets. A simplification of Wronskian or current conserving
theories is proposed which in each order eliminates one integration from the
formulas for higher order corrections. If the PIA is generated by a
non-degenerate eigenvalue of the $\mathbf{R}(x)$ matrix, the eliminated
integration is the only one present. In that case, the simplified theory
becomes fully algorithmic and is generalized to non-hermitian
$\mathbf{R}(x)$ matrices. General theory is illustrated by a few
examples generated automatically by using the author's program in Mathematica
published in arXiv:0710.5406 [math-ph].
\end{abstract}

\maketitle

\section{\label{intr}Introduction}

This paper deals with generalizations of the well known Phase Integral
Approximation.\cite{FandF:book1,as:WKBcp,NF:outline,FandF:dirm,as:piaRMP,%
as:dbpia,as:effintwe,FandF:book2,FandF:book3} This approximation was developed
for solutions of the one-dimensional time independent wave equation,
\begin{equation}
\label{1deq}
u''(x) + R(x) \, u(x) = 0,
\end{equation}
(e.g., $R(x)=\frac{2m}{\hbar^2}[E-V(x)]$ for the Schr\"odinger equation in
cartesian coordinates). Possible generalizations of this theory to sets of ODEs
of similar type:
\begin{equation}
\label{Ndeq}
u_j''(x) + \sum_{k=1}^N R_{jk}(x) \, u_k(x) = 0, \quad j = 1, 2, \dots , N,
\end{equation}
will be described. This can be regarded as going from a ``scalar case'',
Eq.~(\ref{1deq}), to a ``vector case'':
\begin{equation}
\label{vform}
\mathbf{u}''(x) + \mathbf{R}(x) \cdot \mathbf{u}(x) = 0,
\end{equation}
where vector $\mathbf{u}(x)$ and matrix $\mathbf{R}(x)$ have elements
$u_j(x)$ and $R_{jk}(x)$, $j,\,k = 1, 2, \dots , N$ and the dot (here and in
what follows) denotes summation over neighboring indices of vectors and/or
matrices (contraction). Note also that in this paper the convention is adopted
that the prime indicates differentiation with respect to the variable indicated
in the argument of the function in question.

Basic results of scalar theory in a form convenient for generalizations are
described in Sec.~\ref{piasc}. This theory was first generalized to vector cases
with a hermitian positive definite $\mathbf{R}$ matrix in 1979 by S. A. Fulling.
\cite{Full} In Secs.~\ref{genvc} and \ref{Fullht}, Fulling's results will be
presented  in a somewhat modified and more general form and extended to
negative definite $\mathbf{R}(x)$ matrices. The original treatment will also be
commented on briefly. Furthermore, a simplified PIA theory will be proposed,
containing no integrals characteristic for the current or Wronskian conserving
theories.

In lowest order, the phase integral approximation for a two dimensional vector
case ($N=2$) was also introduced independently \cite{eias:BEC} and then
generalized \cite{asei:eigv} to any $N>1$. This theory was useful in providing
initial conditions for numerical integration of the two relevant differential
equations. The eigenvalue problem for these ODEs was solved numerically in the
limit in which the numerical integration interval tended to infinity and the
accuracy required was very high. This calculation would not be possible without
efficient asymptotics at $x \to \infty$ provided by the phase integral
approximation. Extension of this theory (valid also if $\mathbf{R}(x)$ is non
hermitian) to higher orders is possible but turns out to be rather complicated.
A simpler theory of possibly non hermitian vector cases is given in
Sec.~\ref{nonht}.

In Sec~\ref{Neq2}, for the simplest vector case of $N=2$, all earlier discussed
vector theories are compared.

Sec.~\ref{auxfun} describes singularities in the PIA and a possible choice of 
the auxiliary function $a(x)$ present in all earlier theories.

Sec.~\ref{expls} gives examples produced by the author's program in Mathematica
\cite{as:progsM} and Sec.~\ref{concl} contains conclusions.

In the rest of this introductory section, we discuss a few simple facts
pertaining to exact solutions of Eqs.~(\ref{1deq}) and (\ref{vform}).

Equations of the form (\ref{Ndeq}) can be arrived at from somewhat more general
``Schr\"odinger like'' equations:
\begin{equation}
\label{Schle}
\bar{u}_j''(x) + a_j(x) \bar{u}_j'(x) + \sum_{k=1}^N \bar{R}_{jk}(x) \,
\bar{u}_k(x) = 0.
\end{equation}
Using the transformation:
\begin{equation}
\label{ubtr}
u_j(x) = \exp \Bigl[ {\textstyle\frac{1}{2}}\int a_j(x) \, dx \Bigr] \,
\bar{u}_j(x),
\end{equation}
the first derivative terms are eliminated, and equations (\ref{Schle}) take the
form (\ref{Ndeq}) with
\begin{equation}
\label{Rtr}
R_{jk}(x) = \bar{R}_{jk}(x) - \delta_{jk} \, \textstyle{\frac{1}{2}} \,
[ \textstyle{\frac{1}{2}} a_j^2(x) + a_j'(x) ].
\end{equation}
For the radial part of the Schr\"odinger equations in spherical coordinates,
$x = r$ (the spherical radius), and $a_j(r) = 2/r$, leading to
\begin{eqnarray}
u_j(r) &=& r \, \bar{u}_j(r), \quad R_{jk}(r) = \bar{R}_{jk}(r), \quad
R_{jj}(r) = \frac{2m_j}{\hbar^2}[E-V(r)] - \frac{l(l+1)}{r^2}.\label{ubtrS}
\end{eqnarray}

If the function $R(x)$ or the matrix $\mathbf{R}(x)$ is real for real $x$,
as often happens in applications, we can assume that the solution $u(x)$ or
$\mathbf{u}(x)$ is also real. Dealing with complex solutions is either a
question of convenience (e.g., complex exponential solutions for constant
$R(x)$ or $\mathbf{R}(x)$) or is due to the physical meaning of the solution
(e.g., a wave function in quantum mechanics). In any case, however, the real
or imaginary part of a complex solution $u(x)$ or $\mathbf{u}(x)$ is also
a solution.

In the vector case, an important special situation arises if the matrix
$\mathbf{R}(x)$ is hermitian (``hermitian vector case''):
\begin{equation}
\label{hermR}
R_{jk}(x) = R_{kj}^*(x).
\end{equation}
In that case the eigenvalues of $\mathbf{R}(x)$ are real, and the exact solution
of Eq.~(\ref{vform}) conserves the generalized current:\cite{Full}
\begin{eqnarray}
\frac{d}{dx} \, \sigma_N &=& 0, \quad \sigma_N \equiv \text{Im} \sum_{j=1}^N
u_j^*(x) u_j'(x) \equiv \text{Im} \bigl( \mathbf{u}(x),\mathbf{u}'(x) \bigr),
\label{Ndcur}
\end{eqnarray}
where the compact notation is obtained if one introduces the scalar product in
the $N$ dimensional complex Hilbert space ${\mathcal H}^N$, $(\mathbf{a},
\mathbf{b}) \equiv \mathbf{a}^* \!\cdot \mathbf{b} \equiv \sum_{j=1}^N
a_j^* b_j$. For $N=1$, $\sigma_1$ is proportional to the quantum mechanical
probability current $S(= \frac{\hbar}{m}\sigma_1)$.

An important subgroup of hermitian vector cases is the situation where
$\mathbf{R}(x)$ is both hermitian and real (``real hermitian case''),
\begin{equation}
\label{rvcase}
\text{Im} \, R_{jk}(x) = 0 \quad \text{and} \quad   R_{jk}(x) = R_{kj}(x).
\end{equation}
In that case, the eigenvalues of $\mathbf{R}(x)$ are again real but furthermore
we can also assume that the eigenvectors of $\mathbf{R}(x)$ are real.

Note that \emph{both} the hermitian and real hermitian vector cases can be
considered to be generealizations of those important scalar cases in which
$R(x)$ is real. Specialization to real hermitian cases is not only useful for
applications but can often make a more detailed analysis possible.

A direct consequence of the fact that there is no first derivative term in
(\ref{1deq}) is conservation of the  Wronskian $W$:
\begin{equation}
\label{1dW}
\frac{d}{dx} W = 0, \quad W \equiv
\begin{vmatrix}
^{(1)}u(x)&^{(2)}u(x)\\
^{(1)}u'(x)&^{(2)}u'(x)
\end{vmatrix},
\end{equation}
where $^{(1)}u(x)$ and $^{(2)}u(x)$ are two exact solutions of Eq.~(\ref{1deq}).
Eq.~(\ref{1dW}) also holds for complex values of $x$, $R(x)$, ${}^{(1)}u(x)$ and
${}^{(2)}u(x)$.

A possible generalization of the Wronskian $W$ to the vector case is
\begin{equation}
\label{NdWb}
\bar{W}_N \equiv {}^{(1)}\mathbf{u}(x) \!\cdot\! {}^{(2)}\mathbf{u}'(x) -
{}^{(2)}\mathbf{u}(x) \!\cdot\! {}^{(1)}\mathbf{u}'(x).
\end{equation}
The so defined Wronskian is conserved, if ${}^{(1)}\mathbf{u}(x)$ and
${}^{(2)}\mathbf{u}(x)$ are solutions of Eq.~(\ref{vform}) and the matrix
$\mathbf{R}(x)$ is symmetric, again for complex $x$, $\mathbf{R}(x)$,
${}^{(1)}\mathbf{u}(x)$ and ${}^{(2)}\mathbf{u}(x)$.

For our purposes, another conservation rule will be useful, i.e., conservation
of the generalized Wronskian $W_N$ defined as follows, which holds if the matrix
$\mathbf{R}(x)$ is hermitian:
\begin{eqnarray}
\frac{d}{dx} W_N &=& 0, \quad
W_N \equiv  \text{Re} \Bigl[ \bigl({}^{(1)}\mathbf{u}(x),{}^{(2)}\mathbf{u}'(x)
\bigr) - \bigl({}^{(2)}\mathbf{u}(x),{}^{(1)}\mathbf{u}'(x) \bigr) \Bigr],
\label{NdW}
\end{eqnarray}
where $^{(1)}\mathbf{u}(x)$ and $^{(2)}\mathbf{u}(x)$ are (complex or real)
exact solutions of Eq.~(\ref{vform}).

If ${}^{(1)}\mathbf{u}(x)$ and ${}^{(2)}\mathbf{u}(x)$ are two real vector
functions, the two above definitions of the Wronskian are identical, and
the current $\sigma_N$ associated with the complex function
\begin{equation}
\label{upm12}
\mathbf{u}(x) \equiv {}^{(1)}\mathbf{u}(x) + i \: {}^{(2)}\mathbf{u}(x),
\end{equation}
is equal to the Wronskian $W_N$ ($=\bar{W}_N$):
\begin{eqnarray}
\sigma_N &\equiv& \text{Im} \Bigl( {}^{(1)}\mathbf{u}(x) + i \: {}^{(2)}
\mathbf{u}(x), {}^{(1)}\mathbf{u}'(x) + i \: {}^{(2)}\mathbf{u}'(x) \Bigr) =
W_N. \label{sigmpm}
\end{eqnarray}
If these two vector functions are exact solutions of Eq.~(\ref{vform}) for a
real hermitian case, $\mathbf{u}(x)$ is also an exact solution, and the
conserved quantities $\sigma_N$ and $W_N$ are the same. In other words, the
current associated with any complex solution $\mathbf{u}(x)$  for a real
hermitian case, is equal to the Wronskian $W_N$ for $^{(1)}\mathbf{u}(x) =
\text{Re} \: \mathbf{u}(x)$ and $^{(2)}\mathbf{u}(x) = \text{Im} \:
\mathbf{u}(x)$.

In the scalar theory of PIA, it is often convenient or even necessary to go
to the complex $x$ plane. That is the case especially in higher orders, where
the phase integrals are divergent at zeros and certain poles of $R(x)$. If
any such point is located on the real axis, it  must be encircled in the
complex plane. Going to the complex plane, however, in general violates the
hermicity condition (\ref{hermR}) or (\ref{rvcase}). Therefore in the hermitian
theory of PIA, the independent variable will be assumed to be real. To remind
the reader of this requirement, the independent variable in this paper is
denoted by $x$ rather than $z$.

\section{\label{piasc}Phase Integral Approximation in the scalar case}

The Phase Integral Approximation was introduced in 1966 by N. Fr{\"{o}}man
\cite{NF:outline} in an attempt to improve the well known JWKB approximation.
Later it was extended by introducing an arbitrary (base) function which could
make the approximation work at its critical points.
\cite{FandF:dirm,as:piaRMP} Here, this approximation in its most general form
will be rederived so as to make straightforward its generalization to vector
cases. In this section, $x$ and $R(x)$ in Eq.~(\ref{1deq}) are allowed to be
complex.

In our derivation use will be made of a simultaneous transformation of the
independent variable, $x \to x_1$, and the dependent one, $u \to u_1$
(Schwartz transformation), under which Eq.~(\ref{1deq}) conserves its reduced
form:
\begin{equation}
\label{1deq1}
u_1''(x_1) + R_1(x_1) \, u_1(x_1) = 0.
\end{equation}
The transformation $x \to x_1$ introduces the first derivative term in
Eq.~(\ref{1deq}), which can be eliminated by using Eqs.~(\ref{ubtr}) and
(\ref{Rtr}) specialized to $N=1$. The result is
\begin{equation}
\label{Schw}
u_1 = q_1^{1/2}(x) \, u, \quad q_1(x)= \frac{dx_1}{dx}, \quad
R_1(x_1) = q_1^{-2}(x) \, \{R(x) + S_x[q_1]\},
\end{equation}
where $S_x$ is the nonlinear differential operator given by
\begin{eqnarray}
S_x[q] &\equiv& q^{1/2} \, \frac{d^2}{dx^2}  q^{-1/2}
\equiv  {\textstyle\frac{3}{4}} \left[\frac{q'(x)}{q(x)}\right]^2 -
{\textstyle\frac{1}{2}} \frac{q''(x)}{q(x)}
\equiv  {\textstyle\frac{5}{16}}\left[\frac{{q^2\,}'(x)}{q^2(x)}\right]^2 -
{\textstyle\frac{1}{4}} \frac{{q^2\,}''(x)}{q^2(x)}.\label{Sx}
\end{eqnarray}
($S_x[q_1] = -\frac{1}{2} \langle x_1;x\rangle$, where $\langle x_1;x\rangle$
is the Schwartzian derivative. \cite{HeMay}) Note that $S_x[q]$ is single valued
if $q^2(x)$ is, i.e., $q(x)$ either single valued or defined up to its sign.

Three properties of the operator $S_x$ will be used in what follows:\\
Homogeneity:
\begin{equation}
\label{Sxaq}
S_x[\alpha q] = S_x[q] \quad \text{if} \quad \alpha = \text{const}.
\end{equation}
Two rules for differencing the product:
\begin{equation}
\label{Sxprod}
S_x[q_1 \, q_{21}] = S_x[q_1] + \tfrac{1}{8} \dfrac{{q_1^2}'(x)\,
{q_{21}^2}'(x)}{q_1^2\,q_{21}^2} + S_x[q_{21}] = S_x[q_1] + q_1^2
S_{x_1}[q_{21}], \quad x_1 = \int q_1(x) \, dx.
\end{equation}
The first two properties follows immediately from Eq.~(\ref{Sx}), and the third
\cite{HeMay} can be arrived at by considering three Schwartz transformations:\\
$x \to x_1$, as described above,
$x_1 \to x_2$ ($R_1 \to {}^{(1)} \! R_2$) generated by $q_{21} =
\frac{dx_2}{dx_1}$, and $x \to x_2$ ($R \to {}^{(2)} \! R_2$) generated by
$q_{2} = \frac{dx_2}{dx} = q_{2\!1} \, q_1$, and requiring that ${}^{(1)} \!
R_2 \equiv {}^{(2)} \! R_2$.

Eq.~(\ref{1deq1}) can be used as an appropriate analytically solvable model for
Eq.~(\ref{1deq}), in which $R_1(x_1)$ should reflect basic properties of
$R(x)$ in some interval of $x$ under interest. The problem then is to solve
Eq.~(\ref{Schw}) for $q_1(x)$, for given $R(x)$ and $R_1(x_1)$. And with the
proper choice of model, one can expect $q_1(x)$ to be smooth and ``slowly
varying'', and look for convenient approximation schemes. See  Refs.~2, 9 and 11
for the lowest order theory, Ref.~14
for its generalization to higher orders and Refs.~15 and 18
for typical applications.

If one is interested in solving the wave equation (\ref{1deq}) in an adiabatic
region, where the function $R(x)$ changes very little on the characteristic
scale of the solution $u(x)$, the best model seems to be
$R_1(x_1) = c = \text{const}$, e.g., $c=1$. With this choice, $u_1(x_1) =
c^{\pm} \, \exp(\pm ix_1)$, and Eq.~(\ref{Schw}) leads to (we write $q$ instead
of $q_1$):
\begin{equation}
\label{upm}
u(x) = u^{\pm}(x) \equiv c^{\pm} \, q^{-1/2}(x) \exp\Bigl[
\int i\,q(x) \, dx \Bigr],
\end{equation}
where $q(x) \equiv [q^2(x)]^{1/2}$ is double valued (defined up to its sign),
$q^2(x)$ must satisfy
\begin{equation}
\label{qeq}
q^2(x) - S_x[q] = R(x),
\end{equation}
and $c^{\pm}$ are constants.

If $q^2(x)$ is an approximate solution of Eq.~(\ref{qeq}), the two functions
$u^{\pm}(x)$ are also only approximate solutions of Eq.~(\ref{1deq}), called
phase integral approximations. The name evidently appeals to the case of $x$
and $q(x)$ being both real. Another important special case is that of real
$x$ and $q^2(x) < 0$, i.e., $q(x)$ pure imaginary. In these two cases it is
convenient to choose the constants $c^{\pm}$ and the sign of $q(x)$ so that
\begin{equation}
\label{upm1}
u^{\pm}(x) = \lvert q(x) \rvert^{-1/2} \, \begin{cases}
  {\displaystyle\exp\Bigl[ \pm i \int \lvert q(x) \rvert \, dx \Bigr]} &
  \text{if $q^2(x)>0,$}\\
  {\displaystyle\exp\Bigl[ \pm\int \lvert q(x) \rvert \, dx \Bigr]} &
  \text{if $q^2(x)<0.$}
  \end{cases}
\end{equation}
One should realize that the approximations (\ref{upm}) and (\ref{upm1}) (better
or worse) always behave as if they were exact solutions of Eq.~(\ref{1deq}),
e.g., they exactly conserve the current $\sigma_1$  given by Eq.~(\ref{Ndcur}),
and the Wronskian (\ref{1dW}). Thus ($\sigma_1\equiv \text{Im} [u^{\pm \, *}(x)
\, {u^{\pm}}'(x)]$)
\begin{equation}
\label{spia}
\sigma_1^{\pm} =
\begin{cases}
  \pm 1& \text{if $q^2(x)>0$,}\\
  0    & \text{if $q^2(x)<0$,}
  \end{cases} \quad W \equiv
\begin{vmatrix}
  u^{+}(x) & u^{-}(x)\\
  {u^{+}}'(x)& {u^{-}}'(x)
\end{vmatrix} =
\begin{cases}
  -2i& \text{if $q^2(x)>0$,}\\
  -2 & \text{if $q^2(x)<0$,}
\end{cases}
\end{equation}
and $W = \pm i \, 2 c^+ c^-$ in the general case of $u(x)$ given by
Eq.~(\ref{upm}).

If $q^2(x)<0$, current conservation is actually a trivial consequence of the
fact that $u^{\pm}(x)$ are real functions. A non trivial statement, however, is
that the current associated with $u^{+}(x) + i \, u^{-}(x)$ is equal to $- 2$,
i.e., equal to $W$ given by Eq.~(\ref{spia}), in accord with Eq.~(\ref{sigmpm})
for $N=1$.

All these nice features of $u^{\pm}(x)$ are due to the fact that these
approximations \emph{are} exact solutions of some equation of the form
(\ref{1deq}), in which $R(x)$ for given $q^2(x)$ is defined by Eq.~(\ref{qeq}).
This $R(x)$ is single valued if $q^2(x)$ is, and is regular if $q^2(x)$ is
regular and non zero. Furthermore, $R(x)$ is real for real $x$ if $q^2(x)$ is,
which implies current conservation (\ref{spia}). These facts were first pointed
out in Ref.~16.
And as both functions (\ref{upm}) or (\ref{upm1})
are solutions of \emph{the same equation} (\ref{1deq}), each linear combination
of these functions (with real or complex coefficients) will also be a solution.
Therefore it will also conserve the Wronskian $W$ and the current $\sigma_1$.
This property of two exact solutions is by no means obvious for two approximate
solutions, in view of non-linearity of the Wronskian and the current.

In Sec.~\ref{Fullht} the functions $u^{\pm}(x)$ will be generalized for the
vector case to $\mathbf{u}^{\pm}(x)$ so as to conserve the generalized current
(\ref{Ndcur}) in each approximation order.
However, nothing analogous to Eq.~(\ref{1deq}) with $R(x)$ given by
Eq.~(\ref{qeq}), satisfied by $\mathbf{u}^{\pm}(x)$, will exist there.
Nevertheless, linear combinations of $\mathbf{u}^{\pm}(x)$ will be shown to
also conserve the generalized current (\ref{Ndcur}) in successive approximation
orders.

To construct a systematic approximation scheme for $q(x)$ satisfying
Eq.~(\ref{qeq}) we assume that $R(x)$ contains a small parameter $\lambda$:
\begin{equation}
\label{FGa}
R(x) = {\lambda}^{-2} G(x) + a(x), \quad 0 < \lambda \ll 1,
\end{equation}
where $G(x)$ (``greater'' term) represents the dominant contribution to $R(x)$ 
in the $\lambda \to 0$ limit and $a(x)$ is an auxiliary function which can be
chosen in any convenient way. In the scalar case and sometimes also in the
vector case, the small parameter $\lambda$ can be eliminated from the final
results by putting $\lambda = 1$, see Sec.~\ref{auxfun} for more details.

Condition (\ref{FGa}) is a quantitative statement expressing the adiabaticity of
$R(x)$. Indeed, if we freeze $R(x)$ at its value for some $x=x_0$, the
solutions of (\ref{1deq}) will be
\begin{equation}
\label{ux0}
u(x) =
\begin{cases}
\exp ({\displaystyle\pm i2\pi x/L}) & \text{if $R(x_0) > 0$},\\
\exp ({\displaystyle\pm  2\pi x/L}) & \text{if $R(x_0) < 0$},
\end{cases}
\end{equation}
where $L = \lambda 2\pi |G(x_0) + \lambda^2 a(x_0)|^{-1/2}$ is the
characteristic scale for $u(x)$ (the wavelength, or $2\pi$ times the $e$-folding
distance). This scale is small as compared to that for $R(x)$ (which is
$\lambda$ independent). Hence, $L$ can be expected to also be a characteristic
local scale for the exact solutions.

It is convenient to introduce $\lambda$ to Eq.~(\ref{upm}) by replacing
$q \to \lambda^{-1} \, q$. Finally we obtain, in view of $S_x[\lambda^{-1} \,
q] = S_x[q]$, and with an appropriate choice of $c^{\pm}$,
\begin{equation}
\label{upia1}
u = q^{-1/2}(x) \exp\left[i \lambda^{-1} \int q(x) \, dx\right],
\end{equation}
\begin{equation}
\label{qeq1}
G(x) - q^2(x) + \lambda^2 \{ S_x[q] + a(x) \} = 0.
\end{equation}
Denoting $q(x)$ in lowest order by $Q(x)$ we obtain
\begin{equation}
\label{Qsq}
Q^2(x) = G(x).
\end{equation}
This defines two solutions differing in sign, $\pm Q(x)$, which can be
improved in higher orders by adding terms proportional to $\lambda^m$,
$m=1,2,\dots$,
\begin{equation}
\label{sumy}
q(x) = \sum_{m=0} y_m(x) \lambda^m ,
\quad y_0(x) = \pm Q(x).
\end{equation}
Alternatively, we can multiply $\pm Q(x)$ by one plus higher order terms:
\begin{equation}
\label{QY}
q(x) = \pm Q(x) Y(x), \quad Y(x) = \sum_{m=0} Y_m(x) \, \lambda^m , \quad
Y_0(x) \equiv 1.
\end{equation}

The functions $Y_m(x)$ are more convenient to deal with than $y_m(x)$, due to
$Y_0(x) \equiv 1$ (in contrast to $y_0(x) = \pm Q(x) \neq \text{const}$).
First, the applicability condition for the approximation in question if
expressed in terms of $Y_m(x)$ is simply
\begin{equation}
\label{applc}
\lambda^m \, \lvert Y_m(x) \rvert \ll 1, \quad m = 1,2,\ldots.
\end{equation}
And secondly, the recurrence relations for $Y_m(x)$ are simpler than those for
$y_m(x)$. An essential point is that equation for $Y(x)$ which follows from
Eq.~(\ref{qeq1}) is not more and in fact even less complicated than (\ref{qeq1})
if the independent variable is appropriately changed. Using Eq.~(\ref{Sxprod})
we obtain
\begin{equation}
\label{SxQY}
S_x[ (\pm Q) \, Y] = S_x[Q] + Q^2(x)\, S_{\zeta}[Y],
\end{equation}
where
\begin{equation}
\label{zeta}
\zeta = \pm \int Q(x) \, dx.
\end{equation}
Note that while $x$ in Eq.~(\ref{1deq}) is usually a dimensional quantity (e.g.,
the space or time variable), $\zeta$ defined by Eq.~(\ref{zeta}) is
dimensionless. Finally, equation for $Y(x)$ can be written
\begin{equation}
\label{Yeq}
(1 -Y^2)Y^2 + \lambda^2 \Bigl\{ \epsilon_0(x) Y^2 + {\textstyle\frac{3}{4}}
\bigl[ Y'(\zeta) \bigr]^2 - {\textstyle\frac{1}{2}} Y Y''(\zeta) \Bigr\} =
0,
\end{equation}
where the additional term $S_x[Q]$ coming from Eq.~(\ref{SxQY}) has been
incorporated into another dimensionless quantity
\begin{equation}
\label{eps0}
\epsilon_0(x) = \frac{S_x[Q] + a(x)}{Q^2(x)}.
\end{equation}
The corresponding equation for $q(x)$ is
\begin{equation}
\label{qeq2}
(Q^2 - q^2)q^2 + \lambda^2 \Bigl\{ a(x) q^2 + {\textstyle\frac{3}{4}}
\bigl[ q'(x) \bigr]^2 - {\textstyle\frac{1}{2}} q q''(x) \Bigr\} = 0.
\end{equation}

As $\lambda^2$ is the only power of $\lambda$ occurring in Eqs.~(\ref{Yeq}) and
(\ref{qeq2}), $Y$ and $q$ can be expanded in powers of $\lambda^2$ rather than
$\lambda$. Thus, replacing in the expansions (\ref{QY}) or (\ref{sumy}) $m \to
2n$, inserting them into Eq.~(\ref{Yeq}) or (\ref{qeq2}) and equating to zero
the coefficients of $\lambda^{2n}$, we obtain the recurrence relations for
$Y_{2n}$ or $y_{2n}$. Those for $Y_{2n}$ take the simple form ($n \geq 1$):
\begin{equation}
\sum_{\alpha+\beta=n} Y_{2\alpha} Y_{2\beta} -
\sum_{\alpha+\beta+\gamma+\delta=n} Y_{2\alpha} Y_{2\beta} Y_{2\gamma}
Y_{2\delta} + \sum_{\alpha+\beta=n-1} \bigl[ \epsilon_0 Y_{2\alpha} Y_{2\beta} +
{\textstyle\frac{3}{4}} Y_{2\alpha}'(\zeta) Y_{2\beta}'(\zeta)-
{\textstyle\frac{1}{2}} Y_{2\alpha} Y_{2\beta}''(\zeta) \bigr] = 0 .\label{rrel}
\end{equation}
Starting with $Y_0(x) \equiv 1$, one obtains:
\begin{equation}
\label{Y2n}
Y_2(x) = {\textstyle\frac{1}{2}} \epsilon_0, \quad
Y_4(x) = -{\textstyle\frac{1}{8}} \bigl[ \epsilon_0^2 + \epsilon_0''(\zeta)
\bigr],\ldots.
\end{equation}
Eqs.~(\ref{rrel}) and (\ref{Y2n}) were first derived by
N. Fr{\"o}man.\cite{NF:outline}

An explicit form of $Y_{2n}$ defined by Eq.~(\ref{rrel}) is
\begin{eqnarray}
Y_{2n} &=& {\textstyle\frac{1}{2}} \Bigl[ \tilde{\sum_{\alpha+\beta=n}}
Y_{2\alpha} Y_{2\beta} - \tilde{\sum_{\alpha+\beta+\gamma+\delta=n}}
Y_{2\alpha} Y_{2\beta} Y_{2\gamma} Y_{2\delta}\nonumber\\
&&+ \sum_{\alpha+\beta=n-1} \bigl[ \epsilon_0
Y_{2\alpha} Y_{2\beta} + {\textstyle\frac{3}{4}} Y_{2\alpha}'(\zeta)
Y_{2\beta}'(\zeta) - {\textstyle\frac{1}{2}} Y_{2\alpha} Y_{2\beta}''(\zeta)
\bigr] \Bigr],\label{Y2nexpl}
\end{eqnarray}
where the tilde associated with the first two sums means that none of the
subscripts $\alpha, \beta, \gamma, \delta$ in these sums can reach the maximum
value $n$, i.e., $0 \leq \alpha, \beta, \gamma, \delta, \sigma \leq n-1$.

All functions $Y_{2n}(x)$ are polynomials in $\frac{d^p\epsilon_0}{d\zeta^p}$,
$p=0,1,2,\dots$, with rational coefficients, and the relevant formulas through
$Y_{20}(x)$ were first obtained by J. Campbell\cite{Campb} by computer.

General properties of $Y_{2n}(x)$ for arbitrary $n$ are discussed in detail
in Refs.~13 and 16.
Note that all functions $Y_{2n}(x)$ can be expressed in terms of single valued
quantities, $Q^2(x)$, $\epsilon_0(x)$ and derivatives $\frac{d}{dx}$ of these
functions. Furthermore, the relevant formulas contain no complex coefficients,
see the definition of $\epsilon_0(x)$, Eq.~(\ref{eps0}), and the identities
\begin{equation}
\label{derz}
Y_{2\alpha}'(\zeta) Y_{2\beta}'(\zeta) = Q^{-2}(x) Y_{2\alpha}'(x)
Y_{2\beta}'(x), \quad Y_{2\beta}''(\zeta) = Q^{-2}(x) \bigl[ Y_{2\beta}''(x) -
{\textstyle\frac{1}{2}} \, Q^{-2}(x) Q^2\,{}'(x) Y_{2\beta}'(x) \bigr].
\end{equation}
As a consequence, all functions $Y_{2n}(x)$ are invariant under change of
sign of $Q(x)$ and are real if $x$, $Q^2(x)$ and $a(x)$ are real ($Q^2(x)>0$ or
$Q^2(x)<0$). Taking $m=2n$ in the expansion (\ref{QY}) and truncating it at
$n = \mathcal{N}$, we obtain
\begin{equation}
\label{q2Np1}
q(x) = q_{2\mathcal{N}+1}(x) \equiv \pm Q(x) \, \sum_{n = 0}^{\mathcal{N}}
Y_{2n}(x) \, \lambda^{2n}.
\end{equation}
Inserting this $q(x)$ into Eq.~(\ref{upia1}), we obtain two linearly independent
approximate solutions of Eq.~(\ref{1deq}). They are called phase integral
approximations of order $2\mathcal{N}+1$ (as they are related to the JWKB
approximations of this order). We recall that $x$ and $Q^2(x)$ can be complex,
but specialization to real values is useful for applications, see
Eq.~(\ref{upm1}).

The recurrence relations for $y_{n}(x)$ equivalent to (\ref{rrel}) can be
obtained from Eq.~(\ref{qeq2}), if  $a(x)$ is expressed in terms of
$\epsilon_0(x)$ and $Q^2(x)$ by using Eqs.~(\ref{eps0}) and (\ref{Sx}). The
result is
\begin{multline}
Q^2 \sum_{\alpha+\beta=n}
y_{2\alpha} y_{2\beta} -
\sum_{\alpha+\beta+\gamma+\delta=n} y_{2\alpha} y_{2\beta} \, y_{2\gamma} \,
y_{2\delta}\\
+ \sum_{\alpha+\beta=n-1} \Bigl\{ \Bigl( Q^2 \epsilon_0 - S_x[Q] \Bigr)
y_{2\alpha} y_{2\beta} + {\textstyle\frac{3}{4}} y_{2\alpha}'(x)
y_{2\beta}'(x) - {\textstyle\frac{1}{2}} y_{2\alpha} y_{2\beta}''(x) \Bigr\} =
0,\label{rrely}
\end{multline}
which is evidently more complicated than Eq.~(\ref{rrel}). Additional terms with
$Q'(x)$ and $Q''(x)$ present in $S_x[Q]$ are necessary to cancel similar
terms produced in $y_{2n}(x)$ when differencing $Q^2(x)$ present in
Eq.~(\ref{rrely}). Only after these cancellations, the actual dependence of
$y_{2n}(x)$ on $\epsilon_0(\zeta)$ can  emerge $\bigl($rather than separately on
$Q^2(x)$ and $a(x)\bigr)$. This type of dependence is directly seen from
Eqs.~(\ref{Yeq}) and (\ref{rrel})--(\ref{Y2nexpl}) but is rather hard to see
when starting with Eq.~(\ref{qeq2}). Note also that $y_{2n}(x)$ $\bigl(= Q(x)
\, Y_{2n}(x) \bigr)$ is complex if $Q^2(x) < 0$, in contrast to $Y_{2n}(x)$
which is real if $Q^2(x)$ is real.

A relatively simple program in Mathematica \cite{as:progsM} based on
Eq.~(\ref{Y2nexpl}) enables one to generate the corrections $Y_{2n}(x)$,
$n = 1, 2, \ldots$, either in their general form analogous to Eq.~(\ref{Y2n})
(with possible transformation of the derivatives $\tfrac{d}{d\zeta}$ into
$\tfrac{d}{dx}$), or for each given choice of $R(x)$ and $a(x)$.

\section{\label{genvc}General theory of the PIA in vector case}

In the scalar case, the starting point of the phase integral theory are
Eqs.~(\ref{FGa}) and (\ref{upia1}). Their generalization to vector cases is
straightforward, i.e.,
\begin{equation}
\label{FGaN}
\mathbf{R}(x) = \lambda^{-2} \mathbf{G}(x) + a(x) \mathbf{I}, \quad 0 <
\lambda\ll 1,
\end{equation}
\begin{equation}
\label{upiaN}
\mathbf{u}(x) = \mathbf{s}(x) \, q^{-1/2}(x) \, \exp \Bigl[ i \lambda^{-1}
\int q(x) \, dx \Bigr],
\end{equation}
where $\mathbf{I}$ is the unit matrix, $\mathbf{s}(x) \in {\mathcal {H}}^N$ and
$a(x)$ is an auxiliary function. If we choose $a(x) \equiv 0$, Eqs.~(\ref{FGaN})
and (\ref{upiaN}) become equivalent to those proposed by Fulling. \cite{Full}

Inserting Eqs.~(\ref{FGaN}) and (\ref{upiaN}) into Eq.~(\ref{vform}) and
multiplying by $\lambda^2$, we easily find an equation that governs the new
unknowns $\mathbf{s}(x)$ and $q(x)$:
\begin{equation}
\Bigl[ \mathbf{G}(x) - q^2(x)\mathbf{I} \Bigr] \!\cdot\! \mathbf{s}(x) +
\lambda \, 2i q(x) \mathbf{s}'(x) + \lambda^{2} \biggl\{ \mathbf{s}''(x) -
\mathbf{s}'(x) \frac{q'(x)}{q(x)} + \Bigl( S_x[q] + a(x) \Bigr) \mathbf{s}(x)
\biggr\} = 0.\label{qeqN}
\end{equation}

In Fulling's theory, the matrix $\mathbf{R}(x)$ was assumed to be hermitian.
This is equivalent to the hermicity of the matrix  $\mathbf{G}(x)$ now entering
Eq.~(\ref{qeqN}), if we assume that $a(x)\neq 0$ is real for real $x$.

Note that the number of new unknowns ($N+1$) is greater than the number of
Eqs.~(\ref{qeqN}) ($N$). Therefore a constraint upon the unknown vector
$\mathbf{s}(x)$ is needed to guarantee the uniqueness of $\mathbf{u}(x)$.

All theories developed in this paper will start with Eqs.~(\ref{FGaN})--%
(\ref{qeqN}). They will differ in the adopted form of the constraint.

For $N=1$, on replacing $\mathbf{s}(x) \to 1$, $\mathbf{G}(x) \to G(x)$ and
$\mathbf{I} \to 1$, Eq.~(\ref{qeqN}) reduces to (\ref{qeq1}) and
$\mathbf{s}(x)$ is $\lambda$ independent. Therefore in the general case of $N \geq 1$, the expansion of
$\mathbf{s}(x)$ in powers of $\lambda$ must start with the $\lambda$
independent term:
\begin{equation}
\label{ssum}
\mathbf{s}(x) = \sum_{m = 0} \mathbf{s}_m(x) \lambda^m.
\end{equation}

Using Eq.~(\ref{qeqN}) in lowest order ($\lambda^0$) and denoting, as in the
scalar case, the $q(x)$ in lowest order by $Q(x)$, we obtain
\begin{equation}
\label{eigeq}
\Bigl[ \mathbf{G}(x) - Q^2(x)\mathbf{I} \Bigr] \!\cdot \mathbf{s}_0(x) = 0.
\end{equation}
This indicates that $Q^2(x)$ must be an eigenvalue of the matrix
$\mathbf{G}(x)$, and $\mathbf{s}_0(x)$ is the corresponding eigenvector. If
$\mathbf{G}(x)$ is hermitian, the eigenvalue $Q^2(x)$ is real i.e., $Q(x)$ is
either real (if $Q^2(x)>0$) or pure imaginary (if $Q^2(x)<0$). Fulling's paper
\cite{Full} was restricted to $Q^2(x)>0$. Here, as in the scalar case, both
situations will be discussed.

The eigenvalue $Q^2(x)$ can be found as the solution of the characteristic
equation
\begin{equation}
\label{det}
\text{det} \Bigl[ \mathbf{G}(x) - Q^2(x) \mathbf{I} \Bigr] = 0.
\end{equation}
The LHS of Eq.~(\ref{det}) is a polynomial in $Q^2$ of degree $N$.

For reasons explained in Sec.~\ref{piasc}, it is convenient to assume that the
expansion of $q(x)$ in powers of $\lambda$ has the form (\ref{QY}). Repeating
the arguments following Eq.~(\ref{QY}), where the $0$th order equation
(\ref{Qsq}) must now be replaced by Eq.~(\ref{eigeq}), and expressing the
derivatives $\frac{d}{dx}$($=Q \frac{d}{d\zeta}$) in Eq.~(\ref{qeqN}) in terms
of $\frac{d}{d\zeta}$ we obtain
\begin{multline}
Y^2 (Q^{-2} \mathbf{G} - Y^2 \mathbf{I}) \cdot (\mathbf{s} - \mathbf{s}_0) +
(1 -Y^2) Y^2 \mathbf{s}_0
+ \lambda i 2 Y^3 \mathbf{s}'(\zeta)\\
+ \lambda^2 \Bigl\{ Y^2 \mathbf{s}''(\zeta) - Y Y'(\zeta) \mathbf{s}'(\zeta) +
\Bigl[ \epsilon_0 Y^2 + {\textstyle\frac{3}{4}}[Y'(\zeta)]^2 -
{\textstyle\frac{1}{2}} Y Y''(\zeta) \Bigr] \mathbf{s} \Bigr\}= 0,\label{YeqN}
\end{multline}
where $\zeta$ and $\epsilon_0$ are defined by Eqs.~(\ref{zeta}) and
(\ref{eps0}). A distinguishing feature of this equation as compared to that in
the scalar case (\ref{Yeq}) (obtained if we replace $\mathbf{s}, \: \mathbf{s}_0
\to 1$) is the presence of the $\lambda$ term. It contains the pure imaginary
coefficient $i$ and changes sign if $Q$ is changed into $-Q$ (i.e., $d\zeta \to
-d\zeta$). This term is responsible for differences between the vector and
scalar theory.

The LHS of Eq.~(\ref{YeqN}) is a polynomial in $\lambda$ containing only
positive powers $\lambda^m$, $m=1,2,\dots$ (as $\mathbf{s} = \mathbf{s}_0$ and
$Y=1$ in lowest order). Equating to zero the coefficients of $\lambda^m$ we
obtain:
\begin{equation}
\begin{split}
Q^{-2} \sum_{\begin{subarray}{c}
	\alpha+\beta+\sigma=m\\
    \sigma \geq 1\end{subarray}} Y_{\alpha} Y_{\beta} \,
\mathbf{G} \!\cdot \mathbf{s}_{\sigma} -
\sum_{\begin{subarray}{c}
	\alpha+\beta+\gamma+\delta+\sigma=m\\
    \sigma \geq 1\end{subarray}} Y_{\alpha} Y_{\beta} Y_{\gamma} Y_{\delta}
\mathbf{s}_{\sigma} &+ \Bigl( \sum_{\alpha+\beta=m} Y_{\alpha} Y_{\beta}\\
&- \sum_{\substack{
	\alpha+\beta+\gamma+\delta
	=m}} Y_{\alpha} Y_{\beta} Y_{\gamma} Y_{\delta}
\Bigr) \, \mathbf{s}_0 + \sum_{m-1,2} = 0,
\end{split}
\label{mgeq2}
\end{equation}
\begin{eqnarray}
\sum_{m-1,2} &\equiv& 2 \, i \, \sum_{\substack{
	\alpha+\beta+\gamma+\sigma
	=m-1}} Y_{\alpha} Y_{\beta} Y_{\gamma} \mathbf{s}_{\sigma}'(\zeta) +
	\sum_{\substack{
	\alpha+\beta+\sigma
	=m-2}} \Bigl\{ Y_{\alpha} Y_{\beta}
\mathbf{s}_{\sigma}''(\zeta)\nonumber\\
&&- Y_{\alpha} Y_{\beta}'(\zeta)
\mathbf{s}_{\sigma}'(\zeta)+ \Bigl[ \epsilon_0 Y_{\alpha} Y_{\beta} +
{\textstyle\frac{3}{4}} Y_{\alpha}'(\zeta) Y_{\beta}'(\zeta) -
{\textstyle\frac{1}{2}} Y_{\alpha} Y_{\beta}''(\zeta) \Bigr] \mathbf{s}_{\sigma}
\Bigr\}, \label{auxsum} 
\end{eqnarray}
where the second sum on the RHS in Eq.~(\ref{auxsum}) must be dropped if $m=1$.
Writing down explicitly terms in the first four sums in Eq.~(\ref{mgeq2})
corresponding to maximal values of $\alpha, \beta, \gamma, \delta, \sigma$
($= m$), this equation can be written as an implicit form of the recurrence
relations for $Y_m$ and $\mathbf{s}_m$ ($m\geq1$):
\begin{equation}
\label{rrelht}
Y_m \mathbf{s}_0 - {\textstyle \frac{1}{2}} \, Q^{-2} (\mathbf{G} - Q^2
\mathbf{I}) \!\cdot\! \mathbf{s}_m = \mathbf{b}_m,
\end{equation}
where $\mathbf{b}_m$ depends on $Y_{\alpha}$ and $\mathbf{s}_{\sigma}$ with
$\alpha, \sigma \leq m-1$. Using Eqs.~(\ref{mgeq2})--(\ref{rrelht}), we can
easily find the recurrence relations for $\mathbf{b}_m$ ($m\geq2$):
\begin{eqnarray}
\mathbf{b}_m &=& {\textstyle\frac{1}{2}} \Bigl[ \tilde{\sum_{\substack{
	\alpha+\beta+\sigma=m\\
	\sigma \geq 1 }}} Y_{\alpha} Y_{\beta} \bigl( \mathbf{s}_{\sigma} +
	2 (Y_{\sigma}
\mathbf{s}_0 - \mathbf{b}_{\sigma}) \bigr) - \tilde{\sum_{\substack{
	\alpha+\beta+\gamma+\delta+\sigma=m\\
	\sigma \geq 1}}} Y_{\alpha} Y_{\beta} Y_{\gamma} Y_{\delta}
\mathbf{s}_{\sigma}\nonumber\\
&&+ \Bigl( \tilde{\sum_{\alpha+\beta=m}} Y_{\alpha} Y_{\beta} -
\tilde{\sum_{\substack{
	\alpha+\beta+\gamma+\delta=m}}} Y_{\alpha} Y_{\beta} Y_{\gamma} Y_{\delta}
\Bigr) \mathbf{s}_0+ \sum_{m-1,2} \Bigr],\label{mathbf}
\end{eqnarray}
where the tilde associated with the first four sums means that none of the
subscripts $\alpha, \beta, \gamma, \delta, \sigma$ in these sums can reach the
maximum value $m$, i.e., $0 \leq \alpha, \beta, \gamma, \delta, \sigma \leq
m-1$. The process of generating $\mathbf{b}_2, \mathbf{b}_3, \dots$ from
Eq.~(\ref{mathbf}) can be programmed in Mathematica,\cite{as:progsM} and
$\mathbf{b}_1$ can be found by dropping the sums with tildes and with
$\alpha+\beta+\sigma=m-2$. The results are:
\begin{eqnarray}
\mathbf{b}_1 &=&  i \, \mathbf{s}_0'(\zeta),\label{b1}\nonumber\\
\mathbf{b}_2 &=& {\textstyle \frac{1}{2}} ( \epsilon_0 - Y_1^2 ) \,
\mathbf{s}_0 + i \, Y_1 \mathbf{s}_0'(\zeta) + {\textstyle \frac{1}{2}}
\mathbf{s}_0''(\zeta) - Y_1 \mathbf{s}_1 + i \,
\mathbf{s}_1'(\zeta),\nonumber\\
\mathbf{b}_3 &=& - \bigl[ \, Y_1 Y_2 + {\textstyle \frac{1}{4}} Y_1''(\zeta)
\bigr] \mathbf{s}_0 + \bigl[  i \, Y_2 - {\textstyle \frac{1}{2}} Y_1'(\zeta)
\bigr] \mathbf{s}_0'(\zeta) - {\textstyle \frac{1}{2}} (Y_1^2 + 2 Y_2 -
\epsilon_0) \mathbf{s}_1\label{bvm}\\
&&+ i \, Y_1 \mathbf{s}_1'(\zeta) + {\textstyle \frac{1}{2}}
\mathbf{s}_1''(\zeta)- Y_1 \mathbf{s}_2 + i \, \mathbf{s}_2'(\zeta), \ldots ,
\nonumber\\
\mathbf{b}_m &=& \tilde{\mathbf{b}}_m - Y_1 \mathbf{s}_{m-1} + i \,
\mathbf{s}_{m-1}'(\zeta),\nonumber
\end{eqnarray}
($m \ge 2$), where $\tilde{\mathbf{b}}_m$ depends on $Y_\alpha$ and
$\mathbf{s}_{\sigma}$, $\mathbf{s}_{\sigma}'(\zeta)$, \ldots, with $\alpha \leq
m-1$ and $\sigma \leq m-2$ ($\tilde{\mathbf{b}}_m$ is independent of
$\mathbf{s}_{m-1}$, $\mathbf{s}_{m-1}'(\zeta)$,\ldots).

In the original treatment, \cite{Full} Fulling expands $p(x)$ $\bigl(\equiv
q^2(x)\bigr)$ rather than $q(x)$ in powers of $\lambda$ ($=u^{-1}$ in his
notation, $u\gg 1$):
\begin{equation}
\label{psum}
p(x) = q^2(x) = \sum_{m=0} p_m(x) \lambda^m.
\end{equation}
This nonstandard approach introduces unnecessary complication, due to the
appearance of $\sqrt{p(x)}$ $\bigl(= q(x)\bigr)$ in the integrand in
Eq.~(\ref{upiaN}) and in the $\lambda$ term in Eq.~(\ref{qeqN}). As a
consequence, evaluation of the phase integral (\ref{upiaN}) becomes nontrivial
even in the simplest cases, and it would be much more difficult to find
equations analogous to (\ref{rrelht}) and (\ref{mathbf}) with $p_{\alpha}$
instead of $Y_{\alpha}$, valid for any $m$. The functions $p_m(x)$, if needed,
can easily be expressed in terms of $Y_0(x)$ ($\equiv 1$), $Y_1(x)$, \dots,
$Y_m(x)$:
\begin{equation}
\label{pmx}
p_m(x) = Q^2(x) \sum_{\alpha=0}^m Y_{\alpha}(x) Y_{m-\alpha}(x), \quad
m \geq 0.
\end{equation}

In the general discussion that follows, the assumption of $x$ and $a(x)$ being
real and $\mathbf{R}(x)$ hermitian is not necessary, unless a special comment
is made.

An important role in vector theory is played by single valued quantities
(invariant under the replacement $Q \to -Q$, i.e., $d\zeta \to -d\zeta$):
$Q^2(x)$, $\epsilon_0(x)$, $\mathbf{G}(x)$, $\mathbf{s}_0(x)$ and derivatives
$\frac{d}{dx}$ of these functions. Using Eq.~(\ref{rrelht}) along with
(\ref{bvm}) it can be seen that $\mathbf{b}_1$, $Y_1$ and $\mathbf{s}_1$ are
double valued, $\mathbf{b}_2$, $Y_2$ and $\mathbf{s}_2$ are single valued etc.
In general, the even order corrections remain invariant under the replacement
$Q \to -Q$, and the odd order corrections change sign. This means that,
in analogy to the scalar case, the even order corrections can be expressed in
terms of single valued functions $Q^2(x)$, $\epsilon_0(x)$ etc. Using
Eqs.~(\ref{QY}), (\ref{upiaN}) and (\ref{ssum}), we obtain two phase integral
approximations $\mathbf{u}^{\pm}(x)$:
\begin{equation}
\mathbf{u}(x) = \mathbf{u}^{\pm}(x) \equiv c^{\pm} \mathbf{s}^{\pm}(x)
\Bigl[q^{\pm}(x) \Bigr]^{-\frac{1}{2}} \! \exp \Bigl[ i \, \lambda^{-1} \int
q^{\pm}(x) \, dx \Bigr],\label{upmN}
\end{equation}
where $c^{\pm}$ are constants and
\begin{equation}
\label{spm}
\mathbf{s}^{\pm}(x) = \sum_{m =0} \mathbf{s}_m(x) (\pm\lambda)^m, \quad
q^{\pm}(x) = \pm Q(x) \, Y^{\pm}, \quad Y^{\pm} = \sum_{m =0} Y_m(x)
(\pm\lambda)^m.
\end{equation}
Note that the even order contributions to $q^{\pm}(x)$ are double valued (like
those in the scalar case), whereas the odd order contributions (specific to
vector case) are single valued.

Eqs.~(\ref{upmN})--(\ref{spm}) generalize (\ref{upm}) and (\ref{QY}) (with
$m = 2n$) to the vector case. The main difference in comparison to the scalar
case is that $\mathbf{u}^{\pm}(x)$ are not solutions of the \emph{same}
differential equation of the form (\ref{vform}).

In hermitian vector cases, where $Q^2(x)$ is real, we can take 
\begin{equation}
\label{Qpm}
\pm Q(x) =
\begin{cases}
\pm \lvert Q(x) \rvert  & \text{if $Q^2(x)>0$},\\
\mp i \, \lvert Q(x) \rvert  & \text{if $Q^2(x)<0$},
\end{cases}
\end{equation}
and choose the constants $c^{\pm}$ so that $\bigl( {\bar{q}}^{\pm}(x) =
\lvert Q(x) \rvert Y^{\pm}(x) \bigr)$
\begin{equation}
\mathbf{u}^{\pm}(x) = \mathbf{s}^{\pm}(x) \Bigl[ {\bar{q}}^{\pm}(x)
\Bigr]^{-1/2}
\begin{cases}
{\displaystyle \exp\Bigl[ \pm i \lambda^{-1} \int {\bar{q}}^{\pm}(x) \, dx
\Bigr]} & \text{if $Q^2(x)>0,$}\\
{\displaystyle \exp\Bigl[ \pm   \lambda^{-1} \int {\bar{q}}^{\pm}(x) \, dx
\Bigr]} & \text{if $Q^2(x)<0.$}
\end{cases} \label{upmrhc}
\end{equation}
Note that, in general, $\mathbf{s}_0(x)$ is complex. Definite statements
concerning reality etc. can only be made in real hermitian cases in which
$\mathbf{s}_0(x)$ is also real.

Thus in real hermitian cases, if $Q^2(x)>0$ (as in the original Fulling's theory
\cite{Full}), it can easily be seen by inspection that the even order
quantities, $\mathbf{b}_{2n}$, $Y_{2n}$ and $\mathbf{s}_{2n}$, $n = 1,2,\dots$,
are real, and the odd order ones, $\mathbf{b}_{2n-1}$, etc., are pure imaginary,
see Eqs.~(\ref{rrelht})--(\ref{bvm}) in which $d\zeta = \pm \lvert Q(x) \rvert
\, dx$ is real. And if $Q^2(x)<0$, $d\zeta = \mp i\, \lvert Q(x) \rvert \, dx$
becomes pure imaginary which makes both the even and odd order corrections real.
In any case, $\mathbf{s}_{2n+1}(x)$ and $Y_{2n+1}(x)$ (pure imaginary if $Q^2(x) >
0$ and real if $Q^2(x) < 0$) correspond to the upper sign in Eq.~(\ref{Qpm}).
Note that in both cases, the single valued contribution to the integral coming
from $Y_{2n+1}(x)$ is real. This contribution slightly modifies the amplitude of
the phase integral approximation in higher orders. Its role is similar to that
of the factors in the first line in Eq.~(\ref{upmrhc}). The main $x$ dependence
of the approximations $\mathbf{u}^{\pm}(x)$ is given by the contribution to the
integral coming from $Y_{2n}(x)$. The corresponding exponential (like that in
the scalar case) either exhibits strongly oscillatory behavior if
$Q^2(x)>0$, or exponential growth or decay if $Q^2(x)<0$. Note that if
$Q^2(x)>0$, we obtain in analogy to the scalar case:
\begin{equation}
\label{ucc}
\mathbf{u}^-(x) = \bigl[ \mathbf{u}^+(x) \bigr]^*.
\end{equation}
And if $Q^2(x)<0$, the approximations $\mathbf{u}^{\pm}(x)$ are real functions,
again in analogy to the scalar case.

When solving Eq.~(\ref{rrelht}) for $Y_m$ and $\mathbf{s}_m$, an important point
is whether $Q^2(x)$ is a degenerate eigenvalue of $\mathbf{G}$ ($d>1$) or non
degenerate ($d=1$), where $d$ is the dimensionality of the linear subspace
$\mathcal{H}^d$ of eigenvectors $\mathbf{s}_0(x)$ corresponding to the
eigenvalue $Q^2(x)$. Denoting by $\mathcal{H}^{\bot}$ the $N-d$ dimensional
orthogonal complement of $\mathcal{H}^d$ and introducing orthonormal bases: in
$\mathcal{H}^d$, $\{\mathbf{e}_k \}$, $k=1,2 \dots,d$, and in
$\mathcal{H}^{\bot}$$, \{\mathbf{e}_k^{\bot} \}$, $k=1,2,\dots,N-d$, we obtain
\begin{equation}
\label{ebas}
( \mathbf{e}_j, \mathbf{e}_k ) = \delta_{jk}, \quad
\sum_{k=1}^d ( \mathbf{e}_k, \mathbf{s} ) \mathbf{e}_k \equiv \mathbf{s} \quad
\text{if} \quad \mathbf{s} \in \mathcal{H}^d,
\end{equation}
\begin{equation}
\label{eperp}
( \mathbf{e}_j^{\bot}, \mathbf{e}_k^{\bot} ) = \delta_{jk}, \quad
\sum_{k=1}^{N-d} ( \mathbf{e}_k^{\bot}, \mathbf{s} ) \mathbf{e}_k^{\bot} \equiv
\mathbf{s} \quad \text{if} \quad \mathbf{s} \in \mathcal{H}^{\bot}.
\end{equation}
Each vector $\mathbf{s} \in \mathcal{H}^N$ can be decomposed into its component
belonging to $\mathcal{H}^d$, $\text{P}\mathbf{s}$, and that belonging to the
orthogonal complement, $\mathbf{s}^{\bot}$:
\begin{equation}
\label{smdec}
\mathbf{s} = \text{P} \mathbf{s} + \mathbf{s}^{\bot}, \quad
\text{P} \mathbf{s}^{\bot} = 0,
\end{equation}
where P is the orthogonal projection operator, acting in $\mathcal{H}^N$ and
projecting onto $\mathcal{H}^d$:
\begin{equation}
\label{Ps}
\text{P} \mathbf{s} = \sum_{j=1}^d (\mathbf{e}_j, \mathbf{s}) \, \mathbf{e}_j.
\end{equation}
Using Eq.~(\ref{smdec}) for $\mathbf{s} = \mathbf{s}_m$ in Eq.~(\ref{rrelht}),
the contribution to the LHS coming from $\text{P} \mathbf{s}_m$ will be zero,
i.e., we can replace in Eq.~(\ref{rrelht}) $\mathbf{s}_m \to
\mathbf{s}_m^{\bot}$. Multiplying the result by $\mathbf{e}_j^{\bot}$, we obtain
equations governing the coordinates of $\mathbf{s}_m^{\bot}$,
$(\mathbf{e}_k^{\bot}, \mathbf{s}_m^{\bot})$, in the basis
$\{\mathbf{e}_k^{\bot} \}$:
\begin{equation}
\label{Gpeq}
\sum_{k=1}^{N-d} ( G_{jk}^{\bot} - Q^2 \delta_{jk} ) \, (\mathbf{e}_k^{\bot},
\mathbf{s}_m^{\bot}) = - 2 \, Q^2 ( \mathbf{e}_j^{\bot}, \mathbf{b}_m ), \quad
G_{jk}^{\bot} = ( \mathbf{e}_j^{\bot}, \mathbf{G} \cdot \mathbf{e}_k^{\bot} ),
\end{equation}
$j,k=1,2,\dots,N-d$. The linear set of $N-d$ equations (\ref{Gpeq}) is
nonsingular, and its solution defines the orthogonal component
$\mathbf{s}_{m}^{\bot}$:
\begin{equation}
\mathbf{s}_m^{\bot} = -2 Q^2 \sum_{j=1}^{N-d} \mathbf{e}_j^{\bot}
\sum_{k=1}^{N-d} A_{jk}^{\bot} (\mathbf{e}_k^{\bot}, \mathbf{b}_m ), \quad
\mathbf{A}^{\bot} \equiv (\mathbf{G}^{\bot} - Q^2 \mathbf{I}^{\bot})^{-1},
\label{smpex}
\end{equation}
where $\mathbf{G}^{\bot}$ is given Eq.~(\ref{Gpeq}) and $\mathbf{I}^{\bot}$ is
the unit vector in $\mathcal{H}^{\bot}$.

Equations involving coordinates of $\text{P}\mathbf{s}_m$ are obtained on
multiplying Eq.~(\ref{rrelht}) by the basis vectors $\mathbf{e}_k$. It is
convenient to choose the basis $\{\mathbf{e}_k \}$ so that $\mathbf{s}_0$ is
directed along one of the basis vectors, e.g.,
\begin{equation}
\label{s0el}
\mathbf{s}_0 = \lvert \mathbf{s}_0 \rvert \, \mathbf{e}_1.
\end{equation}
Multiplying Eq.~(\ref{rrelht}) by $\mathbf{s}_0$ we obtain an explicit form of
the recurrence relation for $Y_m$:
\begin{equation}
\label{Ymexpl}
Y_m = \lvert \mathbf{s}_0 \rvert^{-2} \bigl[ {\textstyle \frac{1}{2}} \, Q^{-2}
\, \lvert \mathbf{s}_0 \rvert \, (\mathbf{e}_1, \mathbf{G} \cdot
\mathbf{s}_m^{\bot}) + (\mathbf{s}_0, \mathbf{b}_m) \bigr] ,
\end{equation}
and the remaining products lead to
\begin{equation}
\label{knel}
{\textstyle \frac{1}{2}} \, Q^{-2} \, (\mathbf{e}_k, \mathbf{G} \cdot
\mathbf{s}_m^{\bot}) + (\mathbf{e}_k, \mathbf{b}_m) = 0, \quad k > 1.
\end{equation}

Using Eqs.~(\ref{s0el})--(\ref{knel}) for $m=1$ and recalling (\ref{b1}) we
obtain
\begin{equation}
\label{Y1eqgen}
Y_1 = \lvert \mathbf{s}_0 \rvert^{-2} \bigl[ {\textstyle \frac{1}{2}}
\, Q^{-2} \, \lvert \mathbf{s}_0 \rvert \, (\mathbf{e}_1, \mathbf{G} \cdot
\mathbf{s}_1^{\bot}) + i \, Q^{-1} \, \bigl(
\mathbf{s}_0, \mathbf{s}_0'(x) \bigr) \bigr],
\end{equation}
\begin{equation}
\label{ekelpr}
{\textstyle \frac{1}{2}} \, Q^{-2} \, (\mathbf{e}_k, \mathbf{G} \cdot
\mathbf{s}_1^{\bot}) + i \, Q^{-1} \bigl( \mathbf{e}_k, \mathbf{s}_0'(x) \bigr)
= 0, \quad k > 1 .
\end{equation}

The following discussion in the next two sections will be given separately for
the hermitian and non-hermitian $\mathbf{G}$ matrices.

\section{\label{Fullht}Hermitian theories of the PIA}

In this section we assume that the $\mathbf{G}(x)$ matrix is hermitian which
in general requires $x$ to be real. This assumption simplifies the theory as
in that case
\begin{equation}
\label{preq0}
(\mathbf{e}_k, \mathbf{G} \!\cdot\! \mathbf{s}_m^{\bot}) = 0, \quad
k = 1,2,\ldots,d ,
\end{equation}
($\mathbf{G} \cdot \mathbf{e}_k = Q^2 \, \mathbf{e}_k$). Furthermore, the matrix
$\mathbf{G}^{\bot}$ defined by Eq.~(\ref{Gpeq}) is hermitian.

Eqs.~(\ref{Y1eqgen}) and (\ref{ekelpr}) take the form
\begin{equation}
\label{Y1eqht}
Y_1 = i \, Q^{-1} \lvert \mathbf{s}_0 \rvert^{-2} \bigl( \mathbf{s}_0,
\mathbf{s}_0'(x) \bigr),
\end{equation}
\begin{equation}
\label{eks0pr}
\bigl( \mathbf{e}_k, \mathbf{s}_0'(x) \bigr) = 0, \quad k > 1 .
\end{equation}
It can be seen that $\mathbf{s}_0'(x)$ must be orthogonal to all basis vectors
$\mathbf{e}_k$ which are orthogonal to $\mathbf{s}_0$. In view of these $d-1$
requirements one might be tempted to assume that $\mathbf{s}_0'(x)$ is also
orthogonal to $\mathbf{e}_1$,
\begin{equation}
\label{s0s0pr}
( \mathbf{s}_0, \mathbf{s}_0'(x) ) = 0,
\end{equation}
i.e., $\mathbf{s}_0'(x) \in \mathcal{H}^{\bot}$. That is because in that case,
$Y_1$ given by Eq.~(\ref{Y1eqht}) is zero, 
\begin{equation}
\label{Y10ht}
Y_1(x) \equiv 0,
\end{equation}
which means an essential simplification of the theory, see Eqs.~(\ref{bvm}).
And as Eq.~(\ref{s0s0pr}) implies $( \mathbf{s}_0, \mathbf{s}_0 ) =
\text{const}$, then with this assumption, the simplest choice is
\begin{equation}
\label{norms0}
( \mathbf{s}_0(x), \mathbf{s}_0(x) ) \equiv 1 , \quad \text{i.e.,} \quad
\mathbf{s}_0 = \mathbf{e}_1 .
\end{equation}
The constraint
\begin{equation}
\label{elpr0}
( \mathbf{e}_1, \mathbf{e}_1'(x) ) = 0 ,
\end{equation}
is often fulfilled automatically by an orthonormal basis. Otherwise, it
can always be fulfilled by an appropriate choice of the phase factor in
$\mathbf{e}_1$, \cite{Full} i.e., if we take
\begin{equation}
\label{phfact}
\mathbf{e}_1 = \exp(i\,\theta_1) \,  \tilde{\mathbf{e}}_1 \, , \quad
\theta_1 = i \int ( \tilde{\mathbf{e}}_1, \tilde{\mathbf{e}}_1'(x) ) \, dx ,
\end{equation}
where $\theta_1$ is a real quantity, and $\{ \tilde{\mathbf{e}}_k \}$ is an
arbitrary orthonormal basis in $\mathcal{H}^d$ ($\mathbf{e}_k =
\tilde{\mathbf{e}}_k$, if $k > 1$).

Eqs.~(\ref{Y10ht}) and (\ref{pmx}) lead to
\begin{equation}
\label{p1etc}
p_1 = 0, \quad p_2 = 2 Q^2(x) Y_2(x), \quad p_3 = 2 Q^2(x) Y_3(x),\ldots
\end{equation}
Eq.~(\ref{Ymexpl}) for $Y_m$ becomes
\begin{equation}
\label{Ymexhc}
Y_m = ( \mathbf{e}_1, \mathbf{b}_m ),
\end{equation}
where $\mathbf{b}_m$ is given by Eq.~(\ref{bvm}) simplified by $Y_1(x) \equiv
0$.

Note that the RHSs of Eq.~(\ref{rrelht}) with $\mathbf{b}_1$, $\mathbf{b}_2$,
etc. given by Eq.~(\ref{bvm}) with $Y_1(x) \equiv 0$, are simpler than the
analogous expressions given by Fulling, see the Appendix in Ref.~8.
That is because we are expanding $Y$ rather than $q$ in powers of $\lambda$,
where $Y_0=1$ (in contrast to $y_0=Q\neq \text{const}$) and where the $\zeta$
variable is the natural choice. Working with this variable is very convenient
for derivations and presentation of final results.

Replacing in Eqs.~(\ref{s0el})--(\ref{preq0}) $m \to m+1$ and using the last of
Eqs.~(\ref{bvm}) along with (\ref{Y10ht}) and (\ref{norms0}), we obtain
\begin{equation}
\label{Ymp1}
Y_{m+1} = ( \mathbf{e}_1, \tilde{\mathbf{b}}_{m+1} ) + i \, Q^{-1} \,
\bigl( \mathbf{e}_1, \mathbf{s}_m'(x) \bigr),
\end{equation}
\begin{equation}
\label{kgt1}
\bigl( \mathbf{e}_k, \mathbf{s}_m'(x) \bigr) = i \, Q \, ( \mathbf{e}_k,
\tilde{\mathbf{b}}_{m+1} ), \quad k > 1 ,
\end{equation}
where $\tilde{\mathbf{b}}_{m+1}$ depends on $Y_{\alpha}$ with $\alpha \leq m$
and on $\mathbf{s}_{\sigma}$, $\mathbf{s}_{\sigma}'(x)$, etc. but with
$\sigma \leq m-1$. This means that Eq.~(\ref{kgt1}) is the actual and only
constraint upon $\mathbf{s}_m$ that follows from Eq.~(\ref{rrelht}) for given
$m\:(= 1, 2,\ldots)$. Using the decomposition (\ref{smdec}) we can write
\begin{equation}
\label{smprd}
\bigl( \mathbf{e}_k, \mathbf{s}_m'(x) \bigr) =  \Bigl( \mathbf{e}_k,
\frac{d}{dx} \mathbf{s}_m^{\bot} \Bigr) + \Bigl( \mathbf{e}_k, \frac{d}{dx}
\bigl( \text{P} \mathbf{s}_m \bigr) \Bigr),
\end{equation}
for any $k=1,2,\ldots,d$. In order $m$, the first term on the RHS is known, see
Eq.~(\ref{smpex}), and in the second term, the derivative $\tfrac{d}{dx}$ can
be shifted in front of the scalar product if the basis is appropriately chosen.
Indeed, using the definition of the projection operator P, Eq.~(\ref{Ps}),
we can write:
\begin{equation}
\frac{d}{dx} \bigl( \mathbf{e}_k, \text{P} \mathbf{s}_m \bigr) = \Bigl(
\mathbf{e}_k, \frac{d}{dx} \bigl( \text{P} \mathbf{s}_m \bigr) \Bigr) +
\sum_{j=1}^d (\mathbf{e}_j, \mathbf{s}_m) \bigl( \mathbf{e}_k'(x),
\mathbf{e}_j \bigr), \label{diffco}
\end{equation}
where each term in the sum over $j$ in Eq.~(\ref{diffco}) is zero if
$\{\mathbf{e}_k(x)\}$ is the Kato basis,\cite{Full} characterized by
($j,k=1,2\ldots,d$)
\begin{equation}
\label{Kato}
( \mathbf{e}_j(x), \mathbf{e}_k'(x) )=0, \quad \text{i.e.,} \quad \text{P}
\mathbf{e}_k'(x)=0.
\end{equation}
Hence, in the Kato basis (which in particular satisfies our earlier
requirements: (\ref{eks0pr}) with $\mathbf{s}_0 = \mathbf{e}_1$ and
(\ref{elpr0})) we can determine the coordinates $(\mathbf{e}_k, \text{P}
\mathbf{s}_m)$, $k > 1$, by using Eq.~(\ref{kgt1}) and (\ref{diffco}) along
with (\ref{Kato}) in Eq.~(\ref{smprd}) and integrating. The result is
\begin{equation}
(\mathbf{e}_k, \text{P} \mathbf{s}_m) = \int \Bigl( \mathbf{e}_k, \: i \, Q \,
\tilde{\mathbf{b}}_{m+1} - \mathbf{s}^{\bot}_m \, {}'(x)  \Bigr) \, dx, \quad
k > 1.\label{Psm}
\end{equation}
These coordinates are thus defined uniquely by the compatibility condition in
order ($m+1$), Eq.~(\ref{kgt1}). At the same time the coordinate
$(\mathbf{e}_1, \text{P} \mathbf{s}_m)$ measured along $\mathbf{e}_1$ is an
unspecified function. The choice of this function leaves unchanged the remaining
unknowns in order $m$, i.e., $Y_m$ and the coordinates $(\mathbf{e}_k, \text{P}
\mathbf{s}_m)$ measured along $\mathbf{e}_k$, $k=2,3\ldots$. However, this
function will influence $Y_m$ in next order, given by Eq.~(\ref{Ymp1}). Using
again Eqs.~(\ref{smprd}) and (\ref{diffco}) in the Kato basis now for $k=1$,
we obtain
\begin{equation}
\label{Ymp1a}
Y_{m+1} = \Bigl( \mathbf{e}_1, \: \tilde{\mathbf{b}}_{m+1} + i \, Q^{-1}
\frac{d}{dx} \mathbf{s}_m^{\bot} \Bigr) + i \, Q^{-1} \frac{d}{dx}
(\mathbf{e}_1, \text{P} \mathbf{s}_m).
\end{equation}

The choice of the coordinate $(\mathbf{e}_1, \text{P} \mathbf{s}_m)$ must be
compatible with general properties of the even and odd order corrections
discussed in the previous section. This will be the case if we take
\begin{equation}
\label{Pml}
(\mathbf{e}_1, \mathbf{s}_{2n}) = f_{2n}\bigl( Q^2(x), \epsilon_0(x),\ldots
\bigr), \quad (\mathbf{e}_1, \mathbf{s}_{2n-1}) = i \, Q(x) \, f_{2n-1}
\bigl( Q^2(x), \epsilon_0(x),\ldots \bigr),
\end{equation}
$n = 1,2,\ldots$, for any functions $f_{2n}$ and $f_{2n-1}$ of the mentioned
earlier single valued quantities and their derivatives. These functions must be
real for real arguments.

Note that Eq.~(\ref{ebas}) implies
\begin{equation}
\label{Reis0}
\text{Re} \, ( \mathbf{e}_j(x), \mathbf{e}_k'(x) ) = 0.
\end{equation}
Hence if $\text{Im} \, ( \mathbf{e}_j(x), \mathbf{e}_k'(x) )=0$, as is the case
for real eigenvectors, each orthonormal basis (\ref{ebas}) is the Kato basis.
Also in complex hermitian but non degenerate vector cases ($\text{Im} \,
\mathbf{G} \neq 0$, $d=1$) the orthonormal basis (\ref{ebas}) satisfying
Eq.~(\ref{elpr0}) is the Kato basis. Only in complex hermitian and degenerate
vector cases ($\text{Im} \, \mathbf{G} \neq 0$, $d>1$) must one solve the
nonlinear set of 1st order ODEs for the coordinates $e_{jl}(x)$ following from
Eq.~(\ref{Kato})
\begin{equation}
\label{Kato1}
\sum_{l=1}^N e_{jl}^*(x) e_{kl}'(x) = 0, \quad j,k=1,2\ldots,d.
\end{equation}

The results of this section pertaining to hermitian vector cases can be
summarized as follows. Given the orthonormal bases: $\{\mathbf{e}_k \}$ in
$\mathcal{H}^d$ and $\{\mathbf{e}_k^{\bot} \}$ in $\mathcal{H}^{\bot}$, where
$\{\mathbf{e}_k \}$ must be the Kato basis satisfying Eq.~(\ref{Kato}), an
explicit form of the recurrence relation (\ref{rrelht}) is given by
Eqs.~(\ref{smpex}), (\ref{Psm}) and (\ref{Ymexhc}), which define
$\mathbf{s}_m^{\bot}$, $\text{P} \mathbf{s}_m$ and $Y_m$, respectively, in terms
of the same quantities in lower orders. The only freedom in each order is the
choice of the coordinate $(\mathbf{e}_1, \text{P} \mathbf{s}_m)$.
A characteristic feature of the vector theory, in contrast to the scalar one,
is the presence of integrals in recurrence relations. However, they are only
present in degenerate cases ($d > 1$).

We can try to choose the unspecified coordinate $(\mathbf{e}_1, \text{P}
\mathbf{s}_m) \equiv (\mathbf{e}_1, \mathbf{s}_m)$ in each order so as to get
the PIA in the vector case as close as possible to that in the scalar case. We
recall that in the hermitian vector cases, exact solutions of Eq.~(\ref{vform})
conserve both the generalized current $\sigma_N$ and Wronskian $W_N$ given by
Eq.~(\ref{NdW}). In the scalar case, both these quantities are conserved also by
the PIA. In vector cases, the freedom in choice of the coordinate
($\mathbf{e}_1, \mathbf{s}_m$) can be used to conserve one of these quantities.
The second one will be seen not to be conserved. Nevertheless, in real
hermitian cases both the current and the Wronskian can be conserved in view of
Eq.~(\ref{sigmpm}).

If $Q^2(x) > 0$, it is better to conserve the current $\sigma_N$ for each of the
complex PIA $\mathbf{u}^{\pm}(x)$ rather than $W_N$ for these two vector
functions. That is because if this choice is made in a real hermitian case,
two real functions $\text{Re} \, \mathbf{u}^{\pm}(x)$ and  $\text{Im} \,
\mathbf{u}^{\pm}(x)$ (which are approximate solutions of Eq.~(\ref{vform})) will
conserve their Wronskians in view of Eq.~(\ref{sigmpm}). And if $Q^2(x) < 0$, it
is better to conserve $W_N$ for $\mathbf{u}^+(x)$ and $\mathbf{u}^-(x)$ rather
than the currents $\sigma_N$ for each of these approximations.
These currents will not in general be conserved (for complex
$\mathbf{u}^{\pm}(x)$). However,  in a real hermitian case (where the real
approximate solutions $\mathbf{u}^{\pm}(x)$ conserve the current identically),
the current $\sigma_N$ associated with an approximate complex solution
$\mathbf{u}^+(x) + i \, \mathbf{u}^+(x)$ will also be conserved, again in view
of Eq.~(\ref{sigmpm}). The first scenario was put into practice by Fulling
\cite{Full} and the second one will be be described in this paper.

In the scalar case, conservation of the current $\sigma^{\pm}_1$ and Wronskian
$W$ (both involving the products of $u^{\pm}(x)$ and $u^{\pm}{}'(x)$) is due to
the fact that the $x$ dependent factor $|q(x)|$ coming from $u^{\pm}{}'(x)$ is
multiplied by two factors $|q(x)|^{-1/2}$. This mechanism will work also in
hermitian vector cases, see Eq.~(\ref{upmrhc}), if $Y^{\pm}(x)$ is real and
positive. Thus, assuming that $Y^{\pm}(x)>0$ in successive orders (which must be
checked a posteriori) and leaving out the superscripts $\pm$ we obtain
$\bigl(\sigma_N \equiv \text{Im} \, \bigl( \mathbf{u},\mathbf{u}'(x) \bigr)$,
$\bar{q}(x) = |Q(x)| \, Y(x) \bigr)$
\begin{equation}
\sigma_N =
\begin{cases}
\bar{q}^{\,-1}(x) \text{Im} \, \bigl( \mathbf{s},\mathbf{s}'(x) \bigr) \pm
\lambda^{-1} \, ( \mathbf{s},\mathbf{s} ) & \text{if $Q^2(x)>0,$}\\
\bar{q}^{\,-1}(x) \text{Im} \, \bigl( \mathbf{s},\mathbf{s}'(x) \bigr) \,
\exp \bigl[ \pm 2 \lambda^{-1} \int \bar{q}(x) \, dx \bigr] &
\text{if $Q^2(x)<0.$}
\end{cases} \label{sigpia}
\end{equation}
This quantity will be conserved in each order if two constraints proposed by
Fulling \cite{Full} are fulfilled in each order:
\begin{equation}
\label{norm}
( \mathbf{s},\mathbf{s} ) = 1 , \quad \text{normalization,}
\end{equation}
\begin{equation}
\label{sspr0}
( \mathbf{s},\mathbf{s}'(x) ) = 0.
\end{equation}
In fact, the actual constraint is Eq.~(\ref{sspr0}), whereas (\ref{norm}) must
be fulfilled at some fixed value of $x$. That is because  Eq.~(\ref{sspr0})
implies $( \mathbf{s},\mathbf{s} ) = \text{const}$, and so this constraint is
enough to guarantee $\sigma_N = \text{const}$. The second constraint
(\ref{norm}) is rather cosmetic. In this paper it will be fulfilled in lowest
order only.

In lowest order, the constraints (\ref{norm}) and (\ref{sspr0}) reduce to
Eqs.~(\ref{norms0}) and (\ref{elpr0}). In higher orders Eq.~(\ref{sspr0})
takes the form
\begin{equation}
\label{ssprm0}
\sum_{\alpha=0}^m ( \mathbf{s}_{\alpha},\mathbf{s}_{m-\alpha}'(x) ) = 0,
\quad m=1,2\ldots
\end{equation}
(which implies $\sum_{\alpha=0}^m ( \mathbf{s}_{\alpha},\mathbf{s}_{m-\alpha} )
= \text{const}$).

If $Q^2(x) < 0$ and $Y^{\pm}(x)>0$, $W_N$ for $u^{\pm}(x)$ contains the factor
$\exp\bigl[\lambda^{-1} \int |Q(x)| \bigl(Y^+(x) - Y^-(x) \bigr) \, dx
\bigr]$. Conservation of this factor requires that $Y^+(x) \equiv Y^-(x)$, i.e.,
\begin{equation}
\label{Y2np1}
Y_{2n+1}(x) \equiv 0, \quad n = 0,1,\ldots .
\end{equation}
With this condition fulfilled, $W_N$ is given by ($Y = Y^+ = Y^- > 0$)
\begin{equation}
W_N \equiv \bigl[ |Q(x)| Y(x) \bigr]^{-1} \text{Re} \, \Bigl[
\bigl(\mathbf{s}^+, \mathbf{s}^-{}'(x)\bigr) - \bigl(\mathbf{s}^-,
\mathbf{s}^+{}'(x)\bigr) \Bigr] - \lambda^{-1} \, \Bigl[ \bigl(\mathbf{s}^+,
\mathbf{s}^-\bigr) + \bigl(\mathbf{s}^-, \mathbf{s}^+\bigr) \Bigr].\label{ReWN}
\end{equation}
Eq.~(\ref{spm}) implies
\begin{equation}
\label{spmpr}
(\mathbf{s}^{\pm}, \mathbf{s}^{\mp}(x){}') = \sum_{m=0} (\mp\lambda)^m
\sum_{\alpha=0}^m (-1)^{\alpha} (\mathbf{s}_{\alpha},\mathbf{s}_{m-\alpha}'(x)).
\end{equation}
Thus if
\begin{equation}
\label{Wrccnd}
\sum_{\alpha=0}^m (-1)^{\alpha} (\mathbf{s}_{\alpha},\mathbf{s}_{m-\alpha}'(x))
= 0, \quad m = 0, 1, 2, \ldots,
\end{equation}
we obtain
\[
\bigl(\mathbf{s}^+, \mathbf{s}^-{}'(x)\bigr) = \bigl(\mathbf{s}^-,
\mathbf{s}^+{}'(x)\bigr) = 0
\]
in successive orders. This in turn leads to
\[
\bigl[\bigl(\mathbf{s}^+(x), \mathbf{s}^-(x)
\bigr) + \bigl(\mathbf{s}^-(x), \mathbf{s}^+(x)\bigr)\bigr]' = 0. 
\]
Thus, if $Y_{2n}(x)>0$ and conditions (\ref{Y2np1}) and (\ref{Wrccnd}) are
fulfilled, $W_N$ is conserved in successive orders. In lowest order,
Eq.~(\ref{Wrccnd}) is satisfied in view of Eq.~(\ref{elpr0}) ($\mathbf{s}_0 =
\mathbf{e}_1$).

If $Q^2(x) > 0$ and $Y^{\pm}(x)>0$, conservation of $W_N$ for $u^{\pm}(x)$ also
requires $Y_{2n+1}(x)$ to vanish. With this condition fulfilled, $W_N$
takes the form
\begin{equation}
\begin{split}
W_N \equiv \bigl[ \lvert Q(x) \rvert Y(x) \bigr]^{-1} \text{Re} \,\biggl\{ 
&\bigl(\mathbf{s}^+, \mathbf{s}^-{}'(x)\bigr) \exp \Bigl[ - 2 \, \lambda^{-1}
\int \lvert Q(x) \rvert Y(x) \, dx \Bigr]\\
- &\bigl(\mathbf{s}^-, \mathbf{s}^+{}'(x)\bigr) \exp \Bigl[ 2 \, \lambda^{-1}
\int \lvert Q(x) \rvert Y(x) \, dx \Bigr] \biggr\}.
\end{split}\label{WNQ2p}
\end{equation}
This will be zero (and thus conserved) in successive orders, if condition
(\ref{Wrccnd}) is fulfilled.

An important point is that either condition (\ref{ssprm0}) or (\ref{Wrccnd})
with $m>0$ define uniquely the unspecified coordinate $(\mathbf{e}_1,
\mathbf{s}_m)$ (up to the integration constant). The conditions in question can
be written ($S(m=1) \equiv 0$):
\begin{equation}
\label{consc}
\bigl(\mathbf{e}_1, \mathbf{s}_m'(x)\bigr) = - (\pm 1)^m \, \bigl(\mathbf{s}_m,
\mathbf{e}_1'(x) \bigr) - S(m), \quad S(m>1) \equiv \sum_{\alpha=1}^{m-1}
(\pm 1)^{\alpha} \bigl( \mathbf{s}_{\alpha} \mathbf{s}_{m-\alpha}'(x) \bigr).
\end{equation}
Using this result in the identity
\begin{equation}
\label{Pmlpr}
(\mathbf{e}_1(x), \mathbf{s}_m(x))' = ( \mathbf{e}_1'(x),\mathbf{s}_m ) +
( \mathbf{e}_1,\mathbf{s}_m'(x) ),
\end{equation}
and integrating we obtain (see also Eq.~(\ref{Kato}))
\begin{equation}
(\mathbf{e}_1, \mathbf{s}_m) = \int \Bigl[ \bigl(\mathbf{e}_1'(x),
\mathbf{s}_m^{\perp} \bigr) - (\pm 1)^m \bigl( \mathbf{e}_1'(x),
\mathbf{s}_m^{\perp} \bigr)^* - S(m) \Bigr] \, dx.\label{e1smg}
\end{equation}
Grouping together in $S(m=2n)$ and $S(m=2n+1)$, the first and last term, second
and last but one etc., and picking up total derivatives we end up with 
($n=1,2,\ldots$)
\begin{eqnarray}
(\mathbf{e}_1, \mathbf{s}_{2n}) &=& i \, 2 \, \text{Im} \int \Bigl[
\bigl(\mathbf{e}_1'(x), \mathbf{s}_{2n}^{\perp}(x) \bigr) - (\pm 1)^n
\tfrac{1}{2} \bigl(\mathbf{s}_n(x), \mathbf{s}_n'(x) \bigr)\nonumber\\
&&- \sum_{\alpha=1}^{n-1} (\pm 1)^{\alpha} \bigl( \mathbf{s}_{2n-\alpha}(x),
\mathbf{s}_{\alpha}'(x) \bigr) \Bigr] \, dx - (\pm 1)^n \tfrac{1}{2}
\bigl(\mathbf{s}_n(x), \mathbf{s}_n(x) \bigr)\nonumber\\
&&- \sum_{\alpha=1}^{n-1} (\pm 1)^{\alpha} \bigl( \mathbf{s}_{\alpha}(x),
\mathbf{s}_{2n-\alpha}(x) \bigr),\label{e1s2n}
\end{eqnarray}
where the sums over $\alpha$ must be dropped if $n=1$, and
\begin{equation}
(\mathbf{e}_1, \mathbf{s}_{2n+1}) = \int f(x) \, dx - \sum_{\alpha=1}^n
(\pm 1)^{\alpha} \bigl( \mathbf{s}_{\alpha}(x), \mathbf{s}_{2n+1-\alpha}(x)
\bigr),\label{s2np1}
\end{equation}
\begin{equation}
\label{fofx}
f(x) =
\begin{cases}
i 2 \text{Im} \Bigl[ \bigl(\mathbf{e}_1'(x), \mathbf{s}_{2n+1}^{\perp} \bigr) +
{\displaystyle\sum_{\alpha=1}^n}
\bigl( \mathbf{s}_{\alpha}'(x), \mathbf{s}_{2n+1-\alpha} \bigr) \Bigr],\\
2 \text{Re} \Bigl[ \bigl(\mathbf{e}_1'(x), \mathbf{s}_{2n+1}^{\perp} \bigr) +
{\displaystyle\sum_{\alpha=1}^n} (- 1)^{\alpha}
\bigl( \mathbf{s}_{\alpha}'(x), \mathbf{s}_{2n+1-\alpha} \bigr) \Bigr].
\end{cases}
\end{equation}
The last two equations are also valid for $n=0$ if sums over $\alpha$ are
dropped. In that case, the upper line in (\ref{fofx}) gives the Fulling's
result \cite{Full} if we choose $a(x) \equiv 0$. In general, for any $n \geq
0$, the upper line in Eq.~(\ref{fofx}) and upper signs in (\ref{e1s2n})
and (\ref{s2np1}) refer to the current conserving theory  and lower ones to the
Wronskian conserving theory as developed in this paper.

Note that in a real hermitian case, the integrand in Eq.~(\ref{e1s2n}) is real
both if $Q^2(x)>0$ (scalar products of either real or pure imaginary factors)
and if $Q^2(x)<0$ (all factors real). Therefore in that case, $(\mathbf{e}_1,
\mathbf{s}_{2n})$ contains no integral contribution in either current or
Wronskian conserving theory. The integral contribution is absent also in
$(\mathbf{e}_1, \mathbf{s}_{2n+1})$ in the current conserving theory if
$Q^2(x)<0$ and in the Wronskian conserving theory if $Q^2(x)>0$.

In applications, the matrices $\mathbf{R}(x)$ and $\mathbf{G}(x)$ are often real
and symmetric (``real hermitian case''). In that case, if $Q^2(x)>0$ (as in the
original Fulling's theory \cite{Full}), the odd order corrections,
$Y_{2n-1}$ and $\mathbf{s}_{2n-1}$, $n = 1,2,\dots$, are pure imaginary,
see the text following Eq.~(\ref{upmrhc}). In view of Eq.~(\ref{pmx}) the same
is true of $p_{2n-1}$. Confronting this fact with Fulling's statement
\cite{Full} (without proof), that in the hermitian vector case with $Q^2(x)>0$,
all corrections $p_m(x)$ are real, we would obtain $p_{2n-1} \equiv 0$ (i.e.,
$Y_{2n-1} \equiv 0$) for real hermitian cases. We were unable to prove this fact
in general, but the examples given in Sec.~\ref{expls} do confirm this
behavior. Note also, that in real hermitian cases with $Q^2(x)>0$ and
$(\mathbf{e}_1, \mathbf{s}_m)$ given by Eqs.~(\ref{e1s2n})--(\ref{fofx}) (upper
line and signs), not only $\mathbf{u}^{\pm}(x)$ conserve the generalized current
in each order, but that is the case also for linear combinations of
$\mathbf{u}^{\pm}(x)$. Indeed, introducing $\mathbf{w}(x) \equiv c^+
\mathbf{u}^+(x) + c^- \mathbf{u}^-(x)$ and denoting by $\sigma_N$, $\sigma_N^+$
and $\sigma_N^-$ the currents associated with $\mathbf{w}(x)$, $\mathbf{u}^+(x)$
and $\mathbf{u}^-(x)$ we obtain
\begin{equation}
\sigma_N = \lvert c^+ \rvert^2 \sigma_N^+ + \lvert c^- \rvert^2 \sigma_N^-
+ \text{Im} \Bigl[ c^{+\ast} c^- \bigl( \mathbf{u}^+, \mathbf{u}^-{}'(x) \bigr)
+ c^+ c^{-\ast} \bigl( \mathbf{u}^-, \mathbf{u}^+{}'(x) \bigr) \Bigr],
\label{sigNlc}
\end{equation}
where $\text{Im} [\ ] \equiv 0$ in view of Eq.~(\ref{ucc}). At the same time,
the current associated with linear combinations of two PIAs generated by two 
eigenvalues $Q_{(1)}^2(x)$ and $Q_{(2)}^2(x)$ in Fulling's theory in general is
not conserved. For example, for zero order PIAs we obtain (if $Q_{(1)}^2(x) \neq
Q_{(2)}^2(x)$, which implies $(\mathbf{e}_{1\,(1)}, \mathbf{e}_{1\,(2)}) \equiv
0$)
\begin{eqnarray}
\sigma_N &=& \lvert c_{(1)} \rvert^2 \sigma_{N\,(1)} + \lvert c_{(2)} \rvert^2
\sigma_{N\,(2)} + 2 \, \text{Im}
\Bigl[ c_{(1)}^{\ast} f_{(1)}^{\ast}(x) c_{(2)} f_{(2)}(x) \bigl(
\mathbf{e}_{1\,(1)}, \mathbf{e}_{1\,(2)}'(x) \bigr) \Bigr],\label{sigNlca}\\
f_{(1,2)}(x) &=& \lvert Q_{(1,2)}(x) \rvert^{-1/2} \exp \Bigl[ \pm i \,
\int \lvert Q_{(1,2)}(x) \rvert \, dx \Bigr], \label{f12}
\end{eqnarray}
where in general $\text{Im} [\ ] \neq \text{const}$.

Eqs.~(\ref{smpex}), (\ref{Ymp1}) and (\ref{Psm}) along with (\ref{e1s2n})--%
(\ref{fofx}) are the recurrence relations in explicit form, from which the
higher order corrections $\mathbf{s}_m$ and $Y_m$ can be determined
for any $m \geq 1$. They confirm our earlier observation that in the real vector
case (where $\mathbf{e}_j$ can be assumed to be real) and $Q^2(x)>0$, indeed are
the even order corrections $\mathbf{s}_{2\alpha}$ and $Y_{2\alpha}$ real, and
the odd order ones pure imaginary. It can be shown that $Y_2$ ($=
\tfrac{1}{2}\,Q^{-2}\,p_2$) is also real in complex hermitian cases, see
Eq.~(32) in Ref.~8.

For the real hermitian cases with $Q^2(x)<0$, $Q(x)$ becomes pure imaginary, but
all other quantities in the recurrence relations, including $\mathbf{b}_m$ and
$\tilde{\mathbf{b}}_{m+1}$ given by Eqs.~(\ref{bvm}) and $\mathbf{s}_m^{\bot}$
given by  Eq.~(\ref{smpex}), are real. This implies (as mentioned earlier)
reality of the functions $\mathbf{u}^{\pm}(x)$.

Note that the case of $d=N$ (full degeneration) is trivial, as in that case
each vector $\mathbf{u} \in \mathcal{H}^N$ is an eigenvector of the
$\mathbf{G}$ matrix corresponding to the eigenvalue $Q^2$ in question.
This means that $\mathbf{G} \cdot \mathbf{u} = Q^2(x) \, \mathbf{u}$, and
Eqs.~(\ref{vform}) and (\ref{FGaN}) lead to
\begin{equation}
\label{Nscs}
\mathbf{u}''(x) + R(x) \, \mathbf{u} = 0,
\end{equation}
with $R(x) = \lambda^{-2} \, Q^2(x) + a(x)$ ($N$ identical scalar cases).

For $d=N-1$, the orthogonal complement $\mathcal{H}^{\bot}$ is one
dimensional and the orthonormal basis in $\mathcal{H}^{\bot}$ contains only one
vector $\mathbf{e}^{\bot}$. Denoting by $\mathbf{s}^{\bot}$ any vector
belonging to $\mathcal{H}^{\bot}$ we obtain $\mathbf{e}^{\bot} = \lvert
\mathbf{s}^{\bot} \rvert^{-1} \, \mathbf{s}^{\bot}$ and Eq.~(\ref{smpex}) leads
to
\begin{eqnarray}
\mathbf{s}_m^{\bot} &=& - 2 \, Q^2 \, D^{-1} \bigl[ (\mathbf{s}^{\bot},
\tilde{\mathbf{b}}_m) - Y_1 \, (\mathbf{s}^{\bot}, \mathbf{s}_{m-1}) +
i \, Q^{-1} \, (\mathbf{s}^{\bot}, \mathbf{s}_{m-1}'(x) ) \bigr] \,
\mathbf{s}^{\bot},\nonumber\\
D &=& ( \mathbf{s}^{\bot}, \mathbf{G} \cdot \mathbf{s}^{\bot} ) - \lvert
\mathbf{s}^{\bot} \rvert^2 \, Q^2. \label{smp1d}
\end{eqnarray}
(The first two terms in square brackets in (\ref{smp1d}) must be dropped if
$m=1$.) Thus in that case, finding $\mathbf{s}_m^{\bot}$ is very simple, and
the main algebraic problem is to determine the Kato basis in $\mathcal{H}^d$.
Furthermore, in each order one has to find $d$ integrals, see Eqs.~(\ref{Psm})
and (\ref{e1smg})--(\ref{fofx}).

The simplest alternative to calculating $(\mathbf{e}_1,\mathbf{s}_m)$ from
Eqs.~(\ref{e1smg})--(\ref{fofx}) is to choose $f_{2n} = f_{2n-1} \equiv 0$
in Eqs.~(\ref{Pml}), leading to
\begin{equation}
\label{Pml0ht}
(\mathbf{e}_1,\mathbf{s}_m) \equiv 0.
\end{equation}
This simplifies the theory by eliminating an integration in each order. This
choice, which will be referred to as ``simplified hermitian theory'', seems to
be especially attractive in the non-degenerate case in which the eliminated
integration is the only one present. The price, no current or Wronskian
conservation in higher orders, is probably worth paying in view of advantages
of this theory, i.e., its simplicity and absence of logarithmic terms in
the higher order corrections, see Sec.~\ref{expls}.

\section{\label{nonht}Non-hermitian theory of non-degenerate vector cases}

If the $\mathbf{G}$ matrix is non-hermitian, the main complication is non
vanishing of the product in Eq.~(\ref{preq0}). As a consequence,
Eq.~(\ref{ekelpr}) cannot in general be fulfilled. That is why our non-hermitian
theory in this section will only be given for the non-degenerate cases ($d=1$)
in which Eq.~(\ref{ekelpr}) is not present. In this section, the independent
variable $x$ can be complex.

In general, $Y_1$ given by Eq.~(\ref{Y1eqgen}) will be non zero. This will
introduce complication in Eqs.~(\ref{bvm}), (\ref{smpex}) and
(\ref{Ymexpl}). In this connection there is no need for normalization of
$\mathbf{s}_0$ or imposing the differential constraint (\ref{s0s0pr}). Choice
of the multiplication factor (or function of $x$) in the eigenvector should
make the theory as simple as possible. This should be the only criterion also in
those hermitian cases in which the integral in Eq.~(\ref{phfact}) is not
expressible in elementary functions, again leading to $Y_1(x) \neq 0$.

It will be assumed that the only component of $\text{P}\mathbf{s}_m$ which may
be non-vanishing for $d=1$, i.e., that measured along $\mathbf{s}_0$, is zero:
\begin{equation}
\label{Pm1eq0}
( \mathbf{s}_0, \mathbf{s}_m ) \equiv \sum_{j=1}^N s_{0\,j}^* s_{mj} = 0.
\end{equation}
This means that $\mathbf{s}_m$ is equal to $\mathbf{s}_m^{\bot}$, defined by
Eq.~(\ref{smpex}), and $Y_m$ is given by Eq.~(\ref{Ymexpl}), both in terms of
the $N-1$ basis vectors $\mathbf{e}_k^{\bot}$. Dealing with these basis vectors
is inconvenient and can easily be avoided as follows.

Due to our assumption of $d=1$, the rank of the matrix $\mathbf{G} - Q^2
\mathbf{I}$ in Eq.~(\ref{rrelht}) is $N-1$. Assuming that
\begin{equation}
\label{minor}
\begin{vmatrix}
G_{22} - Q^2&G_{23}&\dots&G_{2N}\\
G_{32}&G_{33} - Q^2&\dots&G_{3N}\\
\hdotsfor[2]{4}\\
G_{N2}&G_{N3}&\dots&G_{NN} - Q^2
\end{vmatrix} \neq 0,
\end{equation}
and introducing (denoted by a bar) vectors and matrices in the $N-1$
dimensional Hilbert space $\mathcal{H}^{N-1}$ ($j,k=1,2,\dots,N-1$)
\begin{equation}
\label{defs}
\bar{s}_{mk} = s_{m\:k+1}, \quad \bar{b}_{mk} = b_{m\,k+1}, \quad
\bar{g}(\alpha,1)_k = G_{k+1\,1}, \quad \bar{g}(1,\alpha)_k = G_{1\,k+1}, \quad
\bar{G}_{jk} = G_{j+1\,k+1},
\end{equation}
we can write Eqs.~(\ref{rrelht}) and (\ref{Pm1eq0}) in vector form:
\begin{equation}
\label{vfb}
( G_{11} - Q^2 ) s_{m1} + \bar{\mathbf{g}}(1,\alpha) \cdot \bar{\mathbf{s}}_m =
2 \, Q^2 \, ( Y_m s_{01} - b_{m1} ),
\end{equation}
\begin{equation}
\label{vfa}
( \bar{\mathbf{G}} - Q^2 \bar{\mathbf{I}} ) \cdot \bar{\mathbf{s}}_m = - s_{m1}
\bar{\mathbf{g}}(\alpha,1) + 2 \, Q^2 \, ( Y_m \, \bar{\mathbf{s}}_0 -
\bar{\mathbf{b}}_m ),
\end{equation}
\begin{equation}
\label{vfc}
s_{01}^* \, s_{m1} + (\bar{\mathbf{s}}_0,\bar{\mathbf{s}}_m) = 0.
\end{equation}

Using Eqs.~(\ref{vfb}) multiplied by $s_{01}^*$, (\ref{vfa}) multiplied by
$\lvert s_{01} \rvert^2$  and (\ref{vfc}), we can easily find
\begin{eqnarray}
\bar{\mathbf{s}}_m &=& 2 \, Q^2 \, s_{01}^* \, \bar{\mathbf{D}}^{-1} \cdot
( b_{m1} \bar{\mathbf{s}}_0 - s_{01} \, \bar{\mathbf{b}}_m ),\nonumber\\
\bar{\mathbf{D}} &=& \lvert s_{01} \rvert^2
\, (\bar{\mathbf{G}} - Q^2 \bar{\mathbf{I}}) +(G_{11} - Q^2) \bar{\mathbf{s}}_0
\bar{\mathbf{s}}_0^* - s_{01}^* \, \bar{\mathbf{s}}_0 \,
\bar{\mathbf{g}}(1,\alpha) - s_{01} \,
\bar{\mathbf{g}}(\alpha,1) \bar{\mathbf{s}}_0^*.\label{Dbar}
\end{eqnarray}
The last three terms in $\bar{\mathbf{D}}$ given by Eq.~(\ref{Dbar}) contain
dyadic products of vectors in $\mathcal{H}^{N-1}$, defined as $(\bar{\mathbf{p}}
\bar{\mathbf{q}})_{jk} = \bar{p}_j \bar{q}_k$. Now $s_{m1}$ can be determined
in terms of $\bar{\mathbf{s}}_m$ by using Eq.~(\ref{vfc}),
\begin{equation}
\label{sm1sm}
s_{m1} = - (\bar{\mathbf{s}}_0,\bar{\mathbf{s}}_m/s_{01}^*),
\end{equation}
and $Y_m$ can be found from Eq.~(\ref{vfb}). In the limit $s_{01} \to 0$,
$\bar{\mathbf{s}}_m/s_{01}^*$ in Eq.~(\ref{sm1sm}) is well defined, see
Eq.~(\ref{Dbar}). This will be illustrated by examples given in
Sec.~\ref{expls}.

\section{\label{Neq2}Phase Integral Approximation for $N = 2$}

The case of two equations (\ref{Ndeq}) ($N=2$) is the only situation in which
the algebraic part of either the current or the Wronskian conserving theory
(with $d=1$) in higher orders is very simple. This is because in that case,
both the eigenspace $\mathcal{H}^d$ and its orthogonal complement
$\mathcal{H}^{\bot}$ are one-dimensional. In any of three hermitian
theories and in the non-hermitian one developed in Secs.~\ref{Fullht} and
\ref{nonht}, Eq.~(\ref{det}) is a quadratic in $Q^2$ and its solutions are
\begin{equation}
\label{Qsqq}
Q^2 = {\textstyle \frac{1}{2}}\bigl[ G_{11} + G_{22} \mp \sqrt{\Delta}\bigr],
\quad \Delta = (G_{11} - G_{22})^2 + 4 G_{12}G_{21}.
\end{equation}

In what follows, the double valued odd order corrections $Y_{2n-1}$ and
$\mathbf{s}_{2n-1}$, $n=1,2,\dots$, will be given for the upper sign in
Eq.~(\ref{Qpm}). The sign ambiguity will refer to Eq.~(\ref{Qsqq}).

As a basis eigenvector we can take
\begin{equation}
\mathbf{s}_0(x) \equiv \{ s_{01}(x), \, s_{02}(x) \} = g(x) \, \{ 1 \, , \:
(Q^2 - G_{11})/G_{12} \},\label{ebvec}
\end{equation}
where $g(x)$ can be chosen in any convenient way.

In hermitian vector cases, one should try to make $Y_1(x)$ vanish by
normalizing the above defined $\mathbf{s}_0(x)$ and multiplying it, if
necessary, by the phase factor given by Eq.~(\ref{phfact}), i.e., by taking
$\mathbf{s}_0 = \mathbf{e}_1 \equiv \mathbf{e}$.

In any of the theories under consideration, the basis vector
$\mathbf{s}^{\bot}$ in $\mathcal{H}^{\bot}$ is given by
\begin{equation}
\label{eprp}
\mathbf{s}^{\bot}(x) = \{ - s_{02}^*(x) , \, s_{01}^*(x) \},
\end{equation}
and
\begin{equation}
\mathbf{s}_m^{\bot}(x) = c_m^{\bot}(x) \, {\mathbf{s}}^{\bot}(x), \quad
c_m^{\bot}(x) = - 2 \, Q^2 \, D^{-1} \: (\mathbf{s}^{\bot},\mathbf{b}_m),
\label{smN2}
\end{equation}
where $D$, given either by Eq.~(\ref{smp1d}) for hermitian cases or
Eq.~(\ref{Dbar}) in the non-hermitian theory, takes the form
\begin{equation}
D = \lvert s_{01} \rvert^2 \, G_{22} + \lvert s_{02} \rvert^2 \, G_{11} -
|\mathbf{s}_0|^2 \, Q^2 - \bigl(s_{01}^* \, s_{02} \, G_{12} + s_{01} \,
s_{02}^* \, G_{21}\bigr)
\equiv \pm \lvert \mathbf{s}_0 \rvert^2 \, \sqrt{\Delta}.\label{Den}
\end{equation}
Note that $D$ is independent of the phase factor in $\mathbf{s}_0$.

In Fulling's or Wronskian conserving hermitian theories we obtain
($\mathbf{s}_0 = \mathbf{e}$, $\mathbf{s}^{\bot} = \mathbf{e}^{\bot}$)
\begin{equation}
\mathbf{s}_m(x) = \mathbf{s}_m^{\bot}(x) + c_m(x) \, \mathbf{e},
\label{smFht}
\end{equation}
where $c_m(x) \equiv \bigl(\mathbf{e}(x),\mathbf{s}_m(x)\bigr)$ is given by
Eqs.~(\ref{e1smg})--(\ref{fofx}) (with $\mathbf{e}_1 \to \mathbf{e}$). We
recall that $Y_0(x) \equiv 1$ and $Y_1(x) \equiv 0$. Note also that
$Q^2 D^{-1} $ is real.

In real hermitian cases,
$\mathbf{s}_0$, ${\mathbf{s}}^{\bot}$ and $G_{21} = G_{12}$ are all real, i.e.,
the asterisk can be dropped.

In simplified theories (hermitian or non-hermitian), $\mathbf{s}_m(x)$ is given
by Eq.~(\ref{smFht}) with $c_m(x) \equiv 0$, where $\mathbf{s}^{\bot}(x)$ is
either normalized or non-normalized.

In hermitian theories we obtain, see Eqs.~(\ref{Ymexpl}) and (\ref{preq0}),
\begin{equation}
Y_m(x) = \lvert \mathbf{s}_0 \rvert^{-2} (\mathbf{s}_0,\mathbf{b}_m),
\label{YmN2}
\end{equation}
and in the non-hermitian theory, Eq.~(\ref{vfb}) leads to
\begin{equation}
\label{YmnhN2}
Y_m(x) = \frac{1}{s_{01}} \Bigl[ \frac{Q^{-2} \, c_m^{\bot}(x) \, G_{12} \,
|\mathbf{s}_0|^2}{2 \, s_{01}} + b_{m1} \Bigr].
\end{equation}

In situations where we can eliminate the small parameter $\lambda$ by putting
$\lambda = 1$, $\lvert Y_m(x) \rvert$ should be small as compared to unity, see
Eq.~(\ref{applc}). Furthermore, as in the lowest order $\mathbf{s} =
\mathbf{s}_0$ and $|\mathbf{s}^{\bot}| = |\mathbf{s}_0|$, also the above
defined multipliers $c_m^{\bot}(x)$ and $c_m(x)$ should be small:
\begin{equation}
\label{csmall}
\lvert c_m^{\bot}(x) \rvert, \: \lvert c_m(x) \rvert \ll 1.
\end{equation}
If these requirements are not fulfilled, the phase integral approximation theory
can only be used if we keep the small $\lambda$ parameter in the expansions of
$Y(x)$ and $\mathbf{s}(x)$. Examples of such behavior will be given in
Sec.~\ref{expls}.

\section{\label{auxfun}Singularities in the PIA and choice of the auxiliary
function \lowercase{a}(\lowercase{x})}

For a given number of terms in the expansion (\ref{QY}), the applicability
condition (\ref{applc}) can always be fulfilled in any interval of $x$ if
the functions $Y_m(x)$ are bounded there and we choose $\lambda$ sufficiently
small. If $Y_m(x)$ are not only bounded but small as compared to
unity, we can eliminate the small parmeter $\lambda$ by putting $\lambda = 1$.
The analysis given in this section concerning the choice of auxiliary function
$a(x)$ is pertinent only to such cases. At the same time, the analysis
concerning the singularities of the PIA is valid for any $0 < \lambda \leq 1$.

Putting $\lambda = 1$ means that we treat both terms
on the RHS in Eq.~(\ref{FGaN}) or (\ref{FGa}) on equal footing. In
that case Eq.~(\ref{FGaN}) can be used as the definition of $\mathbf{G}(x)$
or $a(x)$ in terms of $\mathbf{R}(x)$, i.e., for a given set of ODEs.

Denote by $Q_0^2(x)$ an eigenvalue of the matrix $\mathbf{R}(x)$ defining our
set of equations (\ref{vform}). In the scalar case, $Q_0^2(x) = R(x)$.
Eqs.~(\ref{FGaN}) and (\ref{eigeq}) indicate that the matrices $\mathbf{R}(x)$
and $\mathbf{G}(x)$ have the same eigenvector $\mathbf{s}(x)$, and if
$\lambda = 1$, their eigenvalues $Q_0^2(x)$ and $Q^2(x)$ are related by
\begin{equation}
\label{Qsq0a}
Q^2(x) = Q_0^2(x) - a(x).
\end{equation}

In the scalar case, where only even order functions $Y_{2n}(x)$ are present,
the question of accuracy of the PIA is addressed in Refs.~16 and 17. In Ref.~16,
the behavior of $Y_{2n}(x)$ in the vicinity of characteristic points is
examined with particular reference to such points where  $Y_{2n}(x) \to 0$.
If that happens, the PIA tends to an exact solution of the wave equation
(\ref{1deq}). In Ref.~17, very accurate error estimates for the PIA are found,
given in terms of the $\mu$-integral introduced in Ref.~3, and two additional
$\nu$-integrals.\cite{as:effintwe}. Generalization of these results to the
vector case requires a separate treatment. However, as the scalar contributions
to $Y_m(x)$ are always present in the vector case, the points where these
contributions may be large or singular are also pertinent to the vector
case. They correspond to singularities in $\epsilon_0(x)$ and can easily be
determined by using simple models of the functions entering Eq.~(\ref{Qsq0a}).
The power and exponential models discussed in Ref.~16 and 17 are both simple
and useful.

Locating the characteristic point for $Q^2(x)$ (zero or singularity) at $x=0$
and denoting by $Q_{\text{M}}^2(x)$ the model of $Q^2(x)$ in some vicinity of
$x=0$ we can write
\begin{equation}
\label{Qmodel}
Q^2(x) = Q_{\text{M}}^2(x) \Bigl[ 1 + d(x) \Bigr], \quad d(x) \to 0 \quad
\text{as} \quad x \to 0.
\end{equation}
If the small correcting function $d(x)$ satisfies some additional (model
dependent) requirements, one can relate a singular behavior of $\epsilon_0(x)$
at $x=0$ to the behavior of $Q_{\text{M}}^2(x)$. Using then some transformation
$x \to \tilde{x}$ which maps the point $x=0$ into $\tilde{x}=\infty$, we can
also examine possible problems in the PIA at infinity. The simplest
transformation of this sort, $\tilde{x}=1/x$, was shown in Ref.~16 to be the
only transformation (up to translation and scaling) which can conserve
$\epsilon_0(x)$. However, it requires a simultaneous transformation of all
functions entering Eq.~(\ref{Qsq0a}), so as to conserve the products of
$x^2$ times the function in question:
\begin{equation}
\label{QQtil}
x^2 Q^2(x) \equiv \tilde{x}^2  \tilde{Q}^2(\tilde{x}), \quad \text{i.e.,} \quad
\tilde{Q}^2(\tilde{x}) = \tilde{x}^{-4} Q^2(1/\tilde{x}),
\end{equation}
and similarly for $Q_0^2(x)$ and $a(x)$, leading to  $\epsilon_0(x) \equiv
\tilde{\epsilon}_0(\tilde{x})$. Using Eqs.~(\ref{Sx}), (\ref{Sxprod}),
(\ref{eps0})  and (\ref{Qmodel}) we easily find
\begin{equation}
\label{eps0m}
\epsilon_0(x) = \frac{S_x[Q_{\text{M}}] + a(x)}{Q^2(x)}
+ \tfrac{1}{8} \gamma(x) \frac{d'(x) \dfrac{d}{dx} \ln Q_{\text{M}}^2(x)
- 2 d''(x)}{Q^2(x)} + \tfrac{5}{16} \gamma^2(x)
\frac{\bigl[d'(x)\bigr]^2}{Q^2(x)},
\end{equation}
where $\gamma(x)$ is a correction factor close to unity,
\begin{equation}
\label{gcorf}
\gamma(x) = \bigl[1 + d(x)\bigr]^{-1},
\end{equation}
and $Q^2(x)$ is given by Eq.~(\ref{Qmodel}). This $\epsilon_0(x)$ and all three
fractions in Eq.~(\ref{eps0m}) are invariant under the simultaneous
transformation:
\begin{equation}
\label{simtr}
\bigl\{Q_{\text{M}}^2(x), Q_0(x), a(x)\bigr\} \to
\bigl\{ \tilde{Q}_{\text{M}}^2(\tilde{x}), \tilde{Q}_0(\tilde{x}), \tilde{a}
(\tilde{x}) \bigr\} = \tilde{x}^{-4} \bigl\{ Q_{\text{M}}^2(1/\tilde{x}),
Q_0(1/\tilde{x}), a(1/\tilde{x}) \bigr\}, \quad d(x) \to d(1/\tilde{x}).
\end{equation}
In other words, this $\epsilon_0(x)$ and the fractions in question found in the
limit $x \to 0$ are also valid in the limit $\tilde{x} \to \infty$ and vice
versa, if $x$ and $\tilde{x}$ are related by $x \, \tilde{x} = 1$.
Eq.~(\ref{Qmodel}) is also invariant under this transformation.

As both $x$ and $d(x)$ are small, one can expect
$d'(x)$ and $d''(x)$ to be simpler than $d'(\tilde{x})$ and
$d''(\tilde{x})$. If that is the case, the last two fractions in
Eq.~(\ref{eps0m}) should be calculated from this equation rather than its
equivalent for quantities with a tilde. An example of the opposite behavior
is $d(\tilde{x}) = \exp(-\tilde{x})$ as $\tilde{x} \to +\infty$.

The power model is defined by 
\begin{equation}
\label{QMp}
Q_{\text{M}}^2(x) = c x^m, \quad c \neq 0, \quad \text{i.e.,} \quad
\tilde{Q}_{\text{M}}^2
(\tilde{x}) = c \tilde{x}^{\tilde{m}}, \quad \tilde{m} = -(m+4),
\end{equation}
\begin{equation}\label{eps0p}
\epsilon_{00}(x) = \dfrac{\gamma(x)}{16cx^{m+2}}
\Bigl\{ m(m+4) + 2\,\gamma(x)\bigl[ m\,x\,d'(x) - 2\,x^2\,d''(x) \bigr]
+ 5\,\gamma^2(x) \bigl[ x\,d'(x) \bigr]^2 \Bigr\}.
\end{equation}
Here and in what follows, $\epsilon_{00}(x) \equiv \epsilon_0(x)\vert_{a(x)
\equiv0}$ and the discussion will be performed in the $x$ variable.
It can be seen that the first term in braces, which represents the contribution
to $\epsilon_{00}(x)$ coming from $Q_{\text{M}}^2(x)$, is the
leading contribution if $m \neq 0, -4$ and the derivatives of $d(x)$ satisfy
\begin{equation}\label{limdp}
\lim_{x \to 0} \bigl[ x\,d'(x),\:x^2\,d''(x),\ldots \bigr]=0, \quad
\text{equivalent to}
\quad \lim_{\tilde{x} \to \infty} \bigl[ \tilde{x}\,d'(\tilde{x}),\:\tilde{x}^2
\,d''(\tilde{x}),\ldots \bigr]=0.
\end{equation}
In that case, \cite{as:dbpia} the leading contributions to all functions
$Y_{2n}(x)$ in the scalar case (not only to $Y_2(x) = \epsilon_0(x)/2$) come
from $Q_{\text{M}}^2(x)$, often also for $a(x)\neq 0$. Note that
Eqs.~(\ref{Qmodel}) along with (\ref{QMp})--(\ref{limdp}) are valid e.g., if
$Q^2(x)$ is an analytic function which has a zero,
a regular point or a pole at $x=0$ . In that case, $m$ is
an integer, $d(x) = \sum_{k=1}^{\infty} a_k x^k$ and Eq.~(\ref{Qmodel}) is a
Taylor or Laurent expansion of $Q^2(x)$ about $x=0$. In general, condition
(\ref{limdp}) is fulfilled if the derivatives $d'(x)$, $d''(x)$, etc.  are
bounded as $x\to 0$ but it holds also for some singular behavior, e.g., for
$d(x)=x\ln x$, $\sqrt{x}$ etc. Furthermore, the exponent $m$ can be any real
number. All these facts are also valid after the mapping $x \to \tilde{x}$.

The power model says nothing about $\epsilon_{00}(x)$ if
$Q^2(x)$ or $\tilde{Q}^2(\tilde{x})$ is an analytic function which has an
essential singularity at $x=0$ or $\tilde{x}=\infty$. This can be illustrated
by the exponential models, given by
\begin{equation}
\label{QMe1}
Q_{\text{M}}^2(x) = c\,x^{-4}\,\exp(\eta/x), \quad \text{i.e.,} \quad
\tilde{Q}_{\text{M}}^2(\tilde{x}) = c\,\exp(\eta\,\tilde{x}),
\end{equation}
\begin{equation}\label{eps0e1}
\epsilon_{00}(x) = \dfrac{\gamma(x)}{16c\exp(\eta/x)}
\Bigl\{ \eta^2 - 2\,\gamma(x)\bigl[ (4\,x + \eta)\,x^2\,d'(x) + 2\,x^4\,d''(x)
\bigr] + 5\,\gamma^2(x) \bigl[ x^2\,d'(x) \bigr]^2 \Bigr\},
\end{equation}
or
\begin{equation}
\label{QMe2}
Q_{\text{M}}^2(x) = c\,\exp(\eta/x), \quad \text{i.e.,} \quad
\tilde{Q}_{\text{M}}^2(\tilde{x}) = c\,\tilde{x}^{-4}\,\exp(\eta\,\tilde{x}),
\end{equation}
\begin{equation}\label{eps0e2}
\epsilon_{00}(x) = \dfrac{\gamma(x)}{16c\,x^5\,\exp(\eta/x)}
\Bigl\{ \eta\,(\eta - 8\,x) - 2\,\gamma(x)\bigl[ \eta\,x^3\,d'(x) +
2\,x^5\,d''(x) \bigr] + 5\,\gamma^2(x)\,x\,\bigl[ x^2\,d'(x) \bigr]^2 \Bigr\},
\end{equation}
where $\eta$ is a constant phase factor, $\lvert\eta\rvert = 1$. For these
exponential models, condition (\ref{limdp}) can be replaced by a weaker
condition
\begin{equation}\label{limde}
\lim_{x \to 0} \bigl[ x^2\,d'(x),\:x^4\,d''(x),\ldots \bigr]=0, \quad
\text{equivalent to}
\quad \lim_{\tilde{x} \to \infty} \bigl[ d'(\tilde{x}),\:
\,d''(\tilde{x}),\ldots \bigr]=0.
\end{equation}
The RHS of Eq.~(\ref{eps0p}) tends to zero as $x\to 0$ if $m+2 < 0$. That
happens if
\begin{equation}\label{liminf}
\lim_{x\to 0} x^2\,Q^2(x) \equiv \lim_{\tilde{x}\to \infty} 
\tilde{x}^2\,\tilde{Q}^2(\tilde{x}) = \infty,
\end{equation}
i.e., if the limiting values of $\lvert Q^2(x) \rvert$ or $\lvert\tilde{Q}^2
(\tilde{x}) \rvert$ are \textit{large as compared to} $\lvert x \rvert^{-2}$
\textit{or} $\lvert \tilde{x} \rvert^{-2}$. This behavior is favorable for the
PIA in the scalar case ($Y_{2n} \to 0$). In the vector case, everything depends
on the additional terms in $Y_m$ specific to this case, which will be
illustrated by examples given in Sec.~\ref{expls}. For an analytic function,
the condition $m+2 < 0$, equivalent to $\tilde{m}+2 > 0$, is fulfilled at higher
order poles at $x=0$, $m=-3, -4, \ldots$ (e.g., due to strongly singular
potentials at $x=0$ in the radial Schr\"odinger equation) or at simple zeros
($\tilde{m}=-1$), regular points ($\tilde{m}=0$) and poles ($\tilde{m}=1,2,
\ldots$) at $\tilde{x}=\infty$.

If the limits in Eq.~(\ref{liminf}) are either finite or equal to zero, 
i.e., the limiting values of $\lvert Q^2(x) \rvert$ or $\lvert\tilde{Q}^2
(\tilde{x}) \rvert$ are \textit{not large as compared to} $\lvert x\rvert^{-2}$
\textit{or} $\lvert \tilde{x} \rvert^{-2}$, one can expect problems
in the PIA both in the scalar and in the vector case. That happens for
$m+2 \geq 0$ or $\tilde{m}+2 \leq 0$. Thus, in the marginal case of $m=2$
(equivalent to $\tilde{m}=2$), $\epsilon_{00}$ given by Eq.~(\ref{eps0p}) tends
to $-1/(4\,c)$, which can be large if $\lvert c \rvert < 1/4$. In the remaining
cases of $m+2>0$ (i.e., $\tilde{m}+2<0$), the RHS of Eq.~(\ref{eps0p}) tends to
infinity, except for $m=0$ (i.e., $\tilde{m}=-4$). In that special case, in
which $Q(x) \to c \neq 0$ (and the corresponding $\tilde{Q}^2(\tilde{x})$ has
a zero of order four at $\tilde{x}=\infty$),
\begin{equation}\label{epsm0}
\epsilon_{00}(x) = \dfrac{\gamma^2(x)}{16c}
\Bigl\{ 5\,\gamma(x) \bigl[ d'(x) \bigr]^2 - 4\,d''(x) \bigr] \Bigr\}.
\end{equation}
This is finite if the derivatives $d'(x)$ and $d''(x)$ are bounded (e.g., for
analytic functions) but again can be large, especially if $\lvert c \rvert$ is
small. Singular behavior is also possible if the derivatives in question are
singular.

The relevance of condition (\ref{liminf}) to applicability of the PIA
demonstrated here for the power model is in fact quite general. Its necessity,
pertinent both to the scalar and the vector case, can be demonstrated as
follows. If the limits in Eq.~(\ref{liminf}) for $Q^2(x) = Q_0^2(x)$ are either
finite or equal to zero, so that condition (\ref{liminf}) is violated,
$Q_0^2(x)$ can be written
\begin{equation}
\label{Q0sml}
Q_0^2(x) = x^{-2} \, \Bigl[ c_0 + d_0(x) \Bigr], \quad d_0(x) \to 0 \quad
\text{as} \quad x \to 0,
\end{equation}
where the constant $c_0$ may be equal to zero, leading to
\begin{equation}\label{eps0sml}
\epsilon_{00}(x) = - \tfrac{1}{4}\dfrac{1}{c_0 +
d_0(x)} \biggl\{ 1 + \dfrac{1}{c_0 + d_0(x)} \bigl[ x\,d'(x) -
x^2\,d''(x) \bigr] - \tfrac{5}{4} \Bigl[ \dfrac{x\,d'(x)}{c_0 + d_0(x)} \Bigr]^2
\biggr\}.
\end{equation}
This tends to $-1/(4\,c_0)$ if $c_0 \neq 0$, i.e., can be
large if $\lvert c_0 \rvert < 1/4$, and is singular if $c_0 = 0$.

The sufficiency of condition (\ref{liminf}) for applicability of the PIA in the
scalar case is conditional. Any function $Q_0^2(x)$ for which this condition is
fulfilled as $x\to 0$ can be written
\begin{equation}
\label{Q0lrg}
Q_0^2(x) = x^{-2} \, b(x), \quad \lim_{x\to 0} b(x) = \infty,
\end{equation}
leading to
\begin{equation}\label{epslrg}
\epsilon_{00}(x) = - \tfrac{1}{4}\,\bigl[ b(x)
\bigr]^{-1} \, \biggl\{ 1 +  \dfrac{x\,b'(x)}{b(x)} - \tfrac{5}{4} \Bigl[
\dfrac{x\,b'(x)}{b(x)} \Bigr]^2 + \dfrac{x^2\,b''(x)}{b(x)} 
\biggr\}.
\end{equation}
This $\epsilon_{00}(x)$ will tend to zero if $b^{-1}(x)\,
x\,b'(x)/b(x)$ and $b^{-1}(x)\,x^2\,b''(x)/b(x)$ tend to zero as $x\to 0$. That
happens, e.g., for $b(x) = x^m,\:m<0$, and $b(x) = x^m\,\exp(\eta/x)$ as
$x\to 0+$, where $\eta$ is a constant phase factor, $\lvert\eta\rvert=1$,
$\text{Re}\,\eta > 0$ and $m$ is real.

An important point is that if $Q_0^2(x)$ has the form (\ref{Q0sml}),
the behavior of $\epsilon_{00}(x)$ given by
Eq.~(\ref{eps0sml}), unfavorable for the unmodified PIA ($a(x)\equiv 0$),
can always be eliminated by an appropriate choice of the auxiliary function
$a(x)$.

Thus, if we choose the leading term of $a(x)$ to be proportional to
$x^{-2}$,
\begin{equation}
\label{aexp}
a(x) = c_a x^{-2} \bigl[ 1 + d_a(x) \bigr], \quad c_a \neq 0, \quad d_a(x) \to 0
\quad \text{as} \quad x \to 0
\end{equation}
this term will also be the leading term of $Q^2(x)$ given by Eq.~(\ref{Qsq0a}).
This $Q^2(x)$ will be given by Eqs.~(\ref{Qmodel}) and (\ref{QMp}) with
\begin{equation}
\label{Qmodpar}
m = -2, \quad c = c_0 - c_a, \quad d(x) = \dfrac{d_0(x) - c_a\,d_a(x)}{c_0 -
c_a}, \quad c_a \neq c_0.
\end{equation}
Assuming that $d_0(x)$ and $d_a(x)$ satisfy condition (\ref{limdp}), this
condition will also be satisfied by $d(x)$. In that case, using
Eqs.~(\ref{eps0}), (\ref{eps0p}) and (\ref{aexp}) we easily find
\begin{equation}
\label{eps0mod}
\epsilon_0(x) = \Bigl\{ 4\,c\bigl[ 1 + d(x) \bigr] \Bigr\}^{-1}\,
\Bigl[ -1 + 4\,c_a - \gamma(x)\,\bigl[ x\,d'(x) + x^2\,d''(x) \bigr] +
\tfrac{5}{4} \gamma^2(x)\, \bigl[ x\,d'(x) \bigr]^2 + 4\,c_a\,d_a(x) \Bigr].
\end{equation}
Choosing here and in Eq.~(\ref{aexp}) $c_a=1/4$, we obtain $\epsilon_0(x) \to
0$, and Eq.~(\ref{Qsq0a}) gives
\begin{equation}
\label{Qsqopt}
Q^2(x) = Q_0^2(x) - \frac{1}{4 x^2} \bigl[ 1 + d_a(x) \bigr], \quad
\text{i.e.,} \quad \tilde{Q}^2(\tilde{x}) = \tilde{Q}_0^2(\tilde{x}) -
\frac{1}{4 \tilde{x}^2} \bigl[ 1 + d_a(\tilde{x}) \bigr].
\end{equation}
For a finite zero or pole (at $x=0$), Eq.~(\ref{Qsqopt})
is applicable at zeros and the first and second order poles of $Q_0^2(x)$
($m\geq-2$). For an infinite zero or pole ($\tilde{x}\to\infty$), it can be
used at higher order zeros of $\tilde{Q}_0^2(\tilde{x})$ ($\tilde{m}\leq-2$).
In either case, if $m,\tilde{m}=-2$, we must require that $c_0\neq1/4$ in
Eqs.~(\ref{Qmodel}) and (\ref{QMp}) specialized to $Q^2(x) = Q_0^2(x)$. In the
remaining situations, i.e., higher order poles at $x=0$ ($m_0<-2$) and simple
zeros, regular points or poles at $x=\infty$ ($m_0>-2$), the phase integral
approximation with $a(x)\equiv0$ (``non-modified'' approximation) in the scalar
case tends to an exact solution of the wave equation.

One should realize that in the case of a finite zero of $Q_0^2(x)$ (at
$x=0$), the practical usefulness of Eq.~(\ref{Qsqopt}) is limited by the fact
that $Q^2(x)$ will have a simple zero in the close vicinity of $x=0$. This
seldom happens in the remaining situations favorable for modification.

For the Schr\"odinger equation in spherical coordinates, see Eq.~(\ref{ubtrS})
for $N=1$, the modification defined by Eq.~(\ref{Qsqopt}) with $x=r$ and $d_a(x)
\equiv 0$ gives justification to the well known ``Langer modification'',
\cite{FandF:book1} $l(l+1) \to (l + 1/2)^2$.

In vector cases, apart from singularities in higher order corrections due to
singularities in $\epsilon_0(x)$, as discussed above, additional ones occur at
crossing points of the eigenvalues $Q^2(x)$. That is because elements of the
$\mathbf{A}^{\bot}$ matrix given by Eq.~(\ref{smpex}) contain the factor
$D^{-1}$, where $D$ is the determinant of the matrix $\mathbf{G}^{\bot} - Q^2
\mathbf{I}^{\bot}$. This determinant vanishes at the crossing points, thereby
introducing singularities in $\mathbf{s}_m^{\bot}$ given by Eq.~(\ref{smpex}).
For $N=2$, this is directly seen from Eq.~(\ref{Den}), where $\Delta=0$
corresponds to a double zero of the characteristic equation defining $Q^2(x)$.
Denoting by $x_{\text{cr}}$ the crossing point we obtain ($N=2$)
\begin{equation}
\label{Dencr}
D(x) = g(x) \, (x - x_{\text{cr}})^p, \quad g(x_{\text{cr}}) \neq 0,
\end{equation}
where $p=1$. One should expect Eq.~(\ref{Dencr}) to be also valid if $N>2$,
but now with $1 \leq p < N$.

With $D(x)$ given by Eq.~(\ref{Dencr}), $\mathbf{s}_m^{\bot}$ will contain the
factor $(x - x_{\text{cr}})^{-p}$ leading to a singularity at $x =
x_{\text{cr}}$. This factor will be small at sufficiently large distances from
the singularity. At the same time the integral part of the coordinate
$(\mathbf{e}_1, \mathbf{s}_m)$ given by Eqs.~(\ref{e1s2n})--(\ref{fofx}) will
have a factor locally increasing with distance $ \lvert x - x_{\text{cr}}
\rvert$ as $\ln \lvert x - x_{\text{cr}} \rvert$ if $p=1$. This, in contrast to
scalar theory, may lead to situations where the higher order corrections are
never small, unless we keep the small parameter $\lambda$ in the relevant
equations. Note in this connection, that the integrals present in the remaining
coordinates of $\text{P} \mathbf{s}_m$, $(\mathbf{e}_k, \mathbf{s}_m)$,
$k > 1$, will not contain the logarithmic contribution because the integrands in
Eq.~(\ref{Psm}) contain the derivative $\mathbf{s}_m^{\bot}{}'(x)$ rather than
$\mathbf{s}_m^{\bot}(x)$.

\section{\label{expls}Examples}

All examples in this section will be given for $N = 2$, where the algebraic part
of the theory is simple, see Sec.~\ref{Neq2}. In the simplified theories (both
hermitian and  non-hermitian) which are algorithmic for $N = 2$, all corrections
in any order can be determined analytically. This may not be the case for
Fulling's or Wronskian conserving theory due to the integrations present in
Eqs.~(\ref{e1s2n})--(\ref{fofx}). Nevertheless, one can write a symbolic program
in Mathematica pertaining to all theories \cite{as:progsM} which can be useful
in many cases. All results presented in what follows were produced by using this
program. In all examples, we choose  $a(x)\equiv0$ and the eigenvalues $Q^2$ are
real.

\subsection{\label{rhcpos}Simple real hermitian case with $Q^2 > 0$}

The eigenvalue $Q^2$ given by Eq.~(\ref{Qsqq}) takes a simple form if $\Delta$
is a perfect square. An example of this type (real hermitian case) is,
\cite{Full}
\begin{equation}
\label{Fex1}
\mathbf{G}(x) =
\begin{pmatrix} x \, \cos^2 x + \sin^2 x & (x - 1) \cos x \, \sin x\\
                (x - 1) \cos x \, \sin x & x \, \sin^2 x + \cos^2 x
\end{pmatrix},
\quad \Delta = (x - 1)^2,
\end{equation}
where $G_{11} + G_{22} = x + 1$, and so either $Q^2 = 1$ or $Q^2 = x$.

By slightly modifying this $\mathbf{G}$ matrix we can produce simple examples of
other vector cases (real hermitian with $Q^2 < 0$, complex hermitian and
non-hermitian) which will be discussed in the following subsections. Here and in
the following simple examples related to Eq.~(\ref{Fex1}) we assume $x > 0$.
Note that the eigenvalues $Q^2$ are much simpler than the elements of the
$\mathbf{G}$ matrix, which is rather unusual (no complication due to the square
root in Eq.~(\ref{Qsqq})). Another peculiarity of this example is that all
integrals in Eqs.~(\ref{e1s2n})--(\ref{fofx}), as well as those in
Eq.~(\ref{upmrhc}), can be calculated analytically. Using the program in
Mathematica based on Eqs.~(\ref{Qsqq}) -- (\ref{YmnhN2}),
\cite{as:progsM} we can determine analytically all quantities pertaining to any
(reasonable) order of both Fulling's and Wronskian conserving theory. We recall
that in all hermitian theories, the basis can be chosen so that $Y_1(x)
\equiv 0$.

As the eigenvalue $Q^2 = 1$ is $x$ independent, i.e., $\epsilon_0(x) \equiv 0$,
all scalar contributions to $Y_{2n}(x)$ are zero. That means that only terms
specific to the vector case in the higher order corrections $Y_m(x)$ are
present. Both types of contributions are present for $Q^2 = x$. In that case,
we are dealing with a simple zero of $Q^2(x)$ at $x=0$ and the first order pole
at $x=\infty$. The function $\epsilon_0(x) \bigl(= 5/(16 x^3)\bigr)$ has a pole
at $x=0$ (unfavorable behavior at a zero of $Q^2(x)$) and tends to zero as
$x\to\infty$ (favorable behavior at a pole of $Q^2(x)$). For both eigenvalues, 
a singularity should occur at their crossing point, $x=1$.

In Fulling's hermitian theory, $Y_{2n}(x)$ are real as expected and $Y_{2n-1}(x)
\equiv0$ as required for current conservation. (We recall that the Fulling's
hermitian theory requires reality of all corrections $Y_m(x)$, whereas in a real
hermitian case, $Y_{2n-1}(x)$ are pure imaginary.)

If $Q^2 = 1$ we obtain
\begin{equation}
\label{evex1}
\mathbf{e} = \{ \sin x, \, -\cos x\}, \quad \mathbf{e}^{\bot} = \{ \cos x,
\: \sin x\}, \quad \epsilon_0(x) \equiv 0,
\end{equation}
\begin{align}
Y_1(x) &= 0, &c_1^{\bot}(x) &= - \frac{2\,i}{x - 1}, &c_1(x) &= - 4 \,
i \,\ln\lvert x - 1 \rvert,\label{c1ex1}\\
Y_2(x) &= - \, \frac{1}{2} + \frac{2}{x - 1}, &c_2^{\bot}(x) &=
\frac{4}{ (x - 1)^3} - \frac{8\,\ln \lvert x - 1 \rvert}{x - 1},&
c_2(x) &= -\frac{2}{(x - 1)^2} - 8 \, \ln^2 \lvert x - 1 \rvert,\dots.
\label{Y2ex1}
\end{align}
Far away from the singularity ($x \gg 1$), only the multipliers
$c_m^{\bot}(x)$ are small and tend to zero as $x\to\infty$ (the logarithmic
terms are always divided by positive powers of $(x - 1)$). In the same limit,
the corrections $\lvert Y_{2n}(x) \rvert$ are smaller than unity and decrease
with increasing $n$ but tend to non-vanishing constants ($= \tfrac{1}{2},
\tfrac{1}{8}, \tfrac{1}{16}, \tfrac{5}{128}, \ldots$) and the multipliers
$c_m(x)$ are logarithmically divergent. This means that the higher order
corrections in $\mathbf{s}(x)$ for $x\gg 1$ can only be small (in any finite
interval of $x$) if we keep $\lambda$ in the expansions (\ref{QY}) and
(\ref{ssum}) and take it sufficiently small. This seems to be quite a general
feature of Fulling's theory if $Q^2(x) > 0$, see e.g., the results below. 

If $Q^2 = x$ we obtain:
\begin{equation}
\label{evex1p}
\mathbf{e} = \{ \cos x, \: \sin x\}, \quad \mathbf{e}^{\bot} = \{-\sin x, \:
\cos x\}, \quad \epsilon_0(x) = \dfrac{5}{16 x^3},
\end{equation}
\begin{eqnarray}
Y_1(x) &=& 0, \quad c_1^{\bot}(x) = \frac{2 i \sqrt{x}}{x - 1}, \quad
c_1(x) = 4 \, i \Bigl[2 \sqrt{x} - h(x)\Bigr], \quad h(x) = \ln \frac{\sqrt{x}
+ 1}{\lvert \sqrt{x} - 1 \rvert},\label{c1pex1p}\\
Y_2(x) &=& - \frac{2}{x - 1} + \frac{5}{32\,x^3} - \frac{1}{2\,x}, \quad
c_2^{\bot}(x) = \frac{- 32\,x^4 + 64\,x^3 - 29\,x^2 + 6\,x - 1} {2\, x \,
(x - 1)^3} + \frac{8 \, \sqrt{x}}{x - 1} h(x),\nonumber\\
c_2(x) &=& 2\,x \frac{- 16\,x^2 + 32\,x - 17}{ (x - 1)^2 } - 8 h(x)
\Bigl[ h(x) - 4 \sqrt{x} \Bigr],\ldots\label{cor2ex1p}
\end{eqnarray}
The first and third term in $Y_2(x)$ are specific to the vector theory, whereas
the second term ($= \epsilon_0(x)/2)$ is the scalar contribution. All
tend to zero as $x\to\infty$, and this is true for contributions to any
$Y_{2n}(x)$. As for the velocity multipliers, only  $c_1^{\bot}(x)\to 0$ as
$x\to\infty$. In the same limit, $c_2^{\bot}(x)\to c\neq 0$, $\lvert
c_m^{\bot}(x)\rvert \to\infty$ for $n>2$ and $\lvert c_m(x)\rvert \to
\infty$ for any $n$.

In the Wronskian conserving theory, $Y_{2n}(x)$ are again real, $Y_1(x)\equiv0$
and there are no integral contributions to $(\mathbf{e},\mathbf{s}_m)$, as
expected. However, the requirement $Y_{2n-1}(x)\equiv0$ is violated if $n>1$.
Therefore, the Wronskian for $\mathbf{u}^{\pm}(x)$ will be conserved through
second order only. The relevant results are as follows.

If $Q^2 = 1$ we obtain
\begin{eqnarray}
Y_1(x) &\equiv& 0, \quad c_1^{\bot}(x) = - \frac{2\,i}{x - 1}, \quad
c_1(x) = 0,\label{c1ex1w}\\
Y_2(x) &=& - \frac{x + 3}{2 (x - 1)}, \quad c_2^{\bot}(x) = \frac{4}{(x -
1)^3}, \quad c_2(x) = \frac{2}{(x - 1)^2},\label{c2ex1w}\\
Y_3(x) &=& - \frac{2\,i (x + 3)}{(x - 1)^3}, \ldots ,\label{Y3ex1w}
\end{eqnarray}
and the results for $Q^2 = x$ are:
\begin{eqnarray}
Y_1(x) &\equiv& 0, \quad c_1^{\bot}(x) = \frac{2 i \sqrt{x}}{x - 1}, \quad
c_1(x)=0,\label{c1pex1pw}\\
Y_2(x) &=& \frac{2}{x - 1} + \frac{5}{32\,x^3} - \frac{1}{2\,x}, \quad
c_2^{\bot}(x) = \frac{3\,x^2 + 6\,x - 1} {2\, x \, (x - 1)^3}, \quad
c_2(x) = \frac{2\,x}{ (x - 1)^2 },\label{c2ex1pw}\\
Y_3(x) &=& - \frac{2\,i (x + 3)}{(x - 1)^3 \sqrt{x}}, \ldots\label{Y3ex1pw}
\end{eqnarray}

In the simplified hermitian theory, the corrections through 2nd order are
obtained if we put $c_2(x) \equiv 0$ in the above results for the Wronskian
conserving theory. Next order corrections have a similar structure in both
theories but are not identical. The odd order corrections, $Y_3$, $Y_5$, etc.
are non-zero, e.g.,
\begin{equation}
\label{Y3ex1s}
Y_3(x) = - 2 \, i \frac{x + 1}{(x - 1)^3},
\end{equation}
if $Q^2 = 1$, etc. They are all pure imaginary, as expected, and so after
integration ($i \int Y_{2n-1}(x) \, Q(x) \, dx$) will contribute to the
amplitude rather than the phase of $\mathbf{u}^{\pm}(x)$ given by
Eq.~(\ref{upmrhc}).
The behavior of higher order corrections is now fully analogous to that in the
scalar case (tend to zero as $x \to \infty$) except for the (somewhat peculiar)
behavior of $Y_{2n}(x)$ if $Q^2 = 1$, the same as that in Fulling's or Wronskian
conserving theory. The basis vectors $\mathbf{e}$ and $\mathbf{e}^{\bot}$ are
the same in all theories.

\subsection{\label{rhcneg}Simple real hermitian case with $Q^2 < 0$}

By changing sign of the diagonal elements in the $\mathbf{G}$ matrix
(\ref{Fex1}), i.e., for
\begin{equation}
\label{Fex3}
\mathbf{G}(x) =
\begin{pmatrix}
  -(x \, \cos^2 x + \sin^2 x) & (x - 1) \cos x \, \sin x\\
  (x - 1) \cos x \, \sin x & -(x \, \sin^2 x + \cos^2 x)
\end{pmatrix} ,
\end{equation}
the eigenvalues change sign, i.e., either $Q^2 = -1$ or $Q^2 = -x$. The
corrections for the resulting real hermitian case with $Q^2 < 0$ are very
closely related to those for the original real hermitian $\mathbf{G}$ matrix
with $Q^2 > 0$. They are all real and, except for differences in signs and the
absence of $i$ for $Q^2 < 0$, the corrections in the Wronskian or current
conserving theory for $Q^2 < 0$ are the same as those in the current or
Wronskian conserving theory for $Q^2 > 0$, respectively. And in the simplified
hermitian theory, the corrections in both cases are the same in this sense.
However, after dividing the odd order corrections for $Q^2 > 0$ by $i$, one has
to replace
\[
\{e_1,e_2\} \to \{e_1,-e_2\} \quad \text{and} \quad \{e_1^{\bot},e_2^{\bot}\}
\to \{-e_1^{\bot},e_2^{\bot}\}
\]
and change signs in $Y_m$ and $c_m$ for $m \bmod 4 = 1\text{ or }2$ and in
$c_m^{\bot}$ for $m \bmod 4 = 3\text{ or }0$. In particular this means that
the required and actually observed vanishing of $Y_{2n-1}(x)$, $n=1,2,\ldots$,
in Fulling's theory for $Q^2(x)>0$ implies the analogous vanishing of
$Y_{2n-1}(x)$ in the Wronskian conserving theory for $Q^2(x)<0$. In other
words, there should be no problem in using the Wronskian conserving theory
for $Q^2(x)<0$ in contrast to $Q^2(x)>0$, where it can be used through the
second order only.

If $Q^2 = -1$, the results of the  Wronskian conserving theory are, in view of
Eqs.~(\ref{evex1})--(\ref{Y2ex1}),
\begin{equation}
\label{evexB}
\mathbf{e} = \{ \sin x, \, \cos x\}, \quad \mathbf{e}^{\bot} =
\{ -\cos x, \: \sin x\},
\end{equation}
\begin{align}
Y_1(x) &= 0, &c_1^{\bot}(x) &= - \frac{2}{x - 1}, &c_1(x) &= 4
\,\ln\lvert x - 1 \rvert,\label{c1exB}\\
Y_2(x) &= \frac{1}{2} - \frac{2}{x - 1}, &c_2^{\bot}(x) &=
\frac{4}{ (x - 1)^3} - \frac{8\,\ln \lvert x - 1 \rvert}{x - 1},&
c_2(x) &= \frac{2}{(x - 1)^2} + 8 \, \ln^2 \lvert x - 1 \rvert,\dots.
\label{c2exB}
\end{align}
Similarly, the results for $Q^2 = -x$ will be closely related to those given by
Eqs.~(\ref{evex1p})--(\ref{cor2ex1p}), etc.

\subsection{\label{exmp2}Simple complex hermitian cases}

For any hermitian matrix $\mathbf{G}(x)$, on multiplying $G_{12}(x)$ by a
constant phase factor $e^{i \, \varphi}$ and $G_{21}(x)$ by $e^{- i \,
\varphi}$, where $\varphi$ is real, another hermitian $\mathbf{G}$ matrix is
obtained. With this transformation, the eigenvalues $Q^2$, the corrections
$Y_m(x)$ in all theories and the multipliers $c_m \equiv (\mathbf{e}, 
\mathbf{s}_m)$ given by Eqs.~(\ref{e1s2n})--(\ref{fofx}) are left unchanged. In
the remaining results, simple replacements only are needed:
\[
\{e_1,e_2\} \to \{e^{i \, \varphi} \, e_1,e_2\}, \quad
\{e_1^{\bot},e_2^{\bot}\} \to \{e_1^{\bot},e^{- i \, \varphi} \, e_2^{\bot}\},
\quad c_m^{\bot} \to  e^{i \, \varphi} \, c_m^{\bot}.
\]
They leave $\lvert \mathbf{e} \rvert$ and $\lvert \mathbf{e}^{\bot} \rvert$
unchanged.

If one starts with the real hermitian $\mathbf{G}(x)$ matrix (\ref{Fex1}) or
(\ref{Fex3}), one obtains simple examples of complex hermitian cases (e.g., for
$e^{i \, \varphi} = i$ or $(1 + i)/\sqrt{2}$, etc.). Unfortunately, they are not
``complex enough'' so as to avoid making the RHS of Eq.~(\ref{sigpia}) for
$Q^2 < 0$ equal to zero identically in the simplified hermitian theory (which
happens if $\mathbf{s}(x)$ is real). Non-trivial examples require $\varphi$ to
be $x$ dependent. For the simplest choice, $\varphi = x$, the integrand in
Eq.~(\ref{phfact}) is non-zero but the integral is elementary. This means that
the higher order corrections in the simplified hermitian theories (with $Y_1(x)
\equiv 0$) can easily be determined. Unfortunately, the integrals in either
the current or the Wronskian conserving theory,
Eqs.~(\ref{e1smg})--(\ref{fofx}), are non-elementary.

\subsection{\label{nhc}Simple non-hermitian case with $Q^2 > 0$}

On multiplying $G_{12}(x)$ of the $\mathbf{G}$ matrix (\ref{Fex1}) by any real
or pure imaginary number or function and at the same time dividing $G_{21}(x)$
by the same quantity, the eigenvalues $Q^2(x)$ and $\Delta$ given by
Eq.~(\ref{Qsqq})  will be left unchanged. An example of the resulting
non-hermitian matrix is
\begin{equation}
\label{Fex4}
\mathbf{G}(x) =
\begin{pmatrix}
  x \, \cos^2 x + \sin^2 x & 2\,i\,(x - 1) \cos x \, \sin x\\
  - i\,\tfrac{1}{2} \, (x - 1) \cos x \, \sin x & x \, \sin^2 x + \cos^2 x
\end{pmatrix} .
\end{equation}

If $Q^2 = 1$, it is convenient to choose $g(x) = 2 \, \sin x$ in
Eq.~(\ref{ebvec}), as it eliminates the singularity in the basis vectors and 
leads to $Y_1(x) \equiv 0$ (rather unusual in a non-hermitian case). With this
choice we obtain:
\begin{equation}
\label{evex4m}
\mathbf{s}_0 = \{ 2 \, \sin x, \: i\,\cos x\}, \quad \mathbf{s}^{\bot} =
\{ i\,\cos x, \: 2 \, \sin x \}, \quad \epsilon_0(x) \equiv 0.
\end{equation}

The results of our non-hermitian theory are:
\begin{eqnarray}
Y_1(x) &\equiv& 0, \quad c_1^{\bot}(x) = - \frac{8}{ (x - 1) \, d(x) },
\quad d(x) = 5 - 3 \, \cos 2 x,\label{c1pex4m}\\
Y_2(x) &=& \frac{1}{ 4 \, (x - 1)^2 d^2(x) } \bigl[ -59 x^2 + 26 x + 33
+60 (x - 1)^2 \cos 2 x\nonumber\\
&&- 9 (x^2 + 2 x -3) \cos 4 x + 12 (10 \sin 2 x - 3
\sin 4 x) \bigr],\nonumber\\
c_2^{\bot}(x) &=& - 16 \, i \frac{d(x) + 3 (x -  1) \sin 2 x}{ (x - 1)^3
\, d^2(x) }, \dots\label{c2pex4m}
\end{eqnarray}

If $Q^2 = x$, a convenient choice in Eq.~(\ref{ebvec}) is $g(x) = 2 \, \cos x$,
again eliminating the singularity in the basis vectors and leading to $Y_1(x)
\equiv 0$. With this choice we obtain:
\begin{equation}
\label{evex4p}
\mathbf{s}_0 = \{ 2 \, \cos x, \: - i\,\sin x\}, \quad \mathbf{s}^{\bot} =
\{ - i\,\sin x, \: 2 \, \cos x \}, \quad \epsilon_0(x) = \dfrac{5}{16 x^3},
\end{equation}
and the results of the non-hermitian theory are:
\begin{eqnarray}
Y_1(x) &\equiv& 0, \quad c_1^{\bot}(x) = \frac{8 \,\sqrt{x}}{ (x - 1) \, d(x) },
\quad d(x) = 5 + 3 \, \cos 2 x, \label{c1ex4p}\\
Y_2(x) &=& \frac{1}{ 64 \, x^3 \, (x - 1)^2 \, d^2(x) } \bigl[ 528 x^4 + 416 x^3
- 649 x^2 - 590 x + 295 - (960\,x^4 -1920\,x^3\nonumber\\
&&+ 660\,x^2 + 600\,x- 300) \cos 2 x + (432\,x^4 - 288\,x^3 - 99\,x^2\nonumber\\
&&- 90\,x + 45)
\cos 4 x + x^2 \, (x + 1) (960\,\sin 2 x + 288\,\sin 4 x)
\bigr],\nonumber\\
c_2^{\bot}(x) &=& \frac{2\,i}{ x \,(x - 1)^3 \,  d^2(x) } \bigl[ - 15\,x^2 - 30\,x
+ 5 - 3\,(3\,x^2 + 6\,x - 1) \, \cos 2 x 
+ 24\,x^2 (x - 1) \sin 2 x \bigr],\dots\label{c2pex4p}
\end{eqnarray}
Next order corrections can be generated by using an appropriate program
in Mathematica.\cite{as:progsM} All are oscillating and if  $Q^2 = x$,
the amplitude of the oscillations tends to zero as $x\to\infty$. Therefore in
that case, all higher order corrections are small for $x \gg 1$.

Note that in the previous examples pertaining to hermitian cases, only the
basis vectors contained the sin and cos functions and were therefore oscillating
like the the corresponding $\mathbf{G}$ matrices. No such functions were
present in $Y_{2n}(x)$ or the multipliers $c_n^{\bot}(x)$ and $c_n(x)$. In view
of Eqs.~(\ref{Den}) and (\ref{YmN2}), an essential point was that the basis
vectors $\mathbf{e}$ and $\mathbf{e}^{\bot}$ were normalized, in contrast to
those given by Eqs.~(\ref{evex4m}) and (\ref{evex4p}).

\subsection{\label{realrhc}More realistic real hermitian case with $Q^2 < 0$}

When numerically examining small oscillations of a single quantum vortex in
Bose-Einstein condensate,\cite{eias:BEC,asei:eigv} one arrives at the eigenvalue
problem for a set of two one-dimensional Schr\"odinger like differential
equations:
\begin{equation}
\label{uveq}
\begin{split}
&\frac{d^2 u}{d r^2} + \frac{1}{r}\frac{d u}{d r} -
\Bigl[ 2\phi_0^2(r) + \frac{4}{r^2} - 1 + 2\omega + k^2 \Bigr] u - \phi_0^2(r) v
= 0,\\
&\frac{d^2 v}{d r^2} + \frac{1}{r}\frac{d v}{d r} -
\bigl[ 2\phi_0^2(r) - 1 - 2\omega + k^2 \bigr ] v - \phi_0^2(r) u = 0,
\end{split}
\end{equation}
in which $r$ is the cylindrical radius ($0 \leq r < \infty$) $k$ is the
wavenumber and $\omega$ is the frequency of the oscillations ($k,\:\omega >
0$), and $\phi_0(r)$ is the radial profile of the vortex. This profile is a
monotonic function described by a nonlinear 2nd order differential equation and
subject to the boundary conditions $\phi_0(0) = 0$, and $\phi(r) \to 1$ as $r
\to \infty$. To find initial conditions for numerical integration of the
differential equations (\ref{uveq}) at some large but finite value $r =
r_{\text{as}}$, these equations were first transformed to the reduced form
(\ref{vform}) in which $\bigl( u_1(x) = x^{1/2} u(x)$, $u_2(x) = x^{1/2} v(x)$,
$x = r \bigr)$
\begin{equation}
\label{Fex5}
\mathbf{R}(x) =
\begin{pmatrix}
h_0(x) - h_1(x)&h_2(x)\\
h_2(x)&h_0(x) + h_1(x)
\end{pmatrix},
\end{equation}
\begin{align}
h_0(x) & \equiv - 1 - k^2 + d_0(x), \quad
h_1(x)   \equiv 2 (\omega + x^{-2}), \quad
h_2(x)   \equiv -1+d_1(x),\label{h2}\\
d_0(x) & \equiv \frac{1}{4 x^2} + \dfrac{4}{x^4} +
\frac{38}{x^6} + \frac{748}{x^8}, \quad
d_1(x) \equiv \frac{1}{x^2} +\dfrac{2}{x^4} +
\dfrac{19}{x^6} + \dfrac{374}{x^8}, \label{d1}
\end{align}
where Eqs.~(\ref{d1}) follow from the asymptotic expansion of $\phi_0(r)$ as
$r \to \infty$. Choosing $a(x) \equiv 0$ and $\lambda = 1$ (allowed in the
simplified hermitian theory), the eigenvalues of the matrix $\mathbf{G}(x) =
\mathbf{R}(x)$ given by Eq.~(\ref{Fex5}) can be written ($Q^2 < 0$):
\begin{equation}
\lvert Q \rvert = \sqrt{ - h_0(x) \pm r(x)}, \quad
r(x) \equiv \sqrt{h_1^2(x) + h_2^2(x)},\label{rxex5} 
\end{equation}
and the corresponding eigenvectors $\mathbf{s}_0(x)$ are
\begin{equation}
\mathbf{s}_0(x) = g(x) \, \biggl\{ 1, \frac{h_1(x) \mp r(x)}{h_2(x)} \biggr\}
\label{s0ex5},
\end{equation}
where the factor $g(x)$ can be chosen in any convenient way.
In Refs.~10 and 19,
$g(x) \equiv 1$, leading to
\begin{equation}
Y_1(x) = \mp \gamma_1(x) \frac{ r(x) \mp h_1(x) }{ 2 r^2(x) h_2(x) |Q(x)|},
\quad c_1^{\bot}(x) = \pm \gamma_1(x) \frac{ \lvert Q(x) \rvert}{2 r^3(x)},\quad
\gamma_1(x) \equiv h_1(x) h_2'(x) - h_2(x) h_1'(x).\label{g1ex5}
\end{equation}
Second order corrections are much more complicated.

It can easily be shown that $c_1^{\bot}(x)$ is independent of the factor $g(x)$
in Eq.~(\ref{s0ex5}) and thus is given by Eq.~(\ref{g1ex5}) for any $g(x)$. In
contrast, $Y_1(x)$ depends on the choice of $g(x)$. In particular we obtain
$Y_1(x) \equiv 0$ if $\mathbf{s}_0(x)$ is normalized.

In Refs.~10 and 19,
an approximate analytical solution of Eq.~(\ref{vform}) tending to zero as $x
\to \infty$ was  looked for in the form of a linear combination of two zero
order phase integral approximations $\mathbf{u}^-(x)$, see Eq.~(\ref{upmrhc})
($Q^2(x) < 0$ and $\lambda = 1$). These approximations were referred to as
$\mathbf{u}_{\text{ge}}$ (``greater exponent'', for the upper sign in
Eq.~(\ref{rxex5}) with $h_0(x) < 0$) and $\mathbf{u}_{\text{se}}$ (``smaller
exponent''):
\begin{equation}
\mathbf{u}(x) = C_{\text{se}} \, \mathbf{u}_{\text{se}}(x) + C_{\text{ge}} \,
\mathbf{u}_{\text{ge}}(x).\label{lincmb}
\end{equation}
Eq.~(\ref{lincmb}) was used for $x \geq x_{\text{as}}$, with $x_{\text{as}}$
determined experimentally as the minimal value above which the eigenvalue
$\omega(k)$ was practically independent of $x_{\text{as}}$. We were interested
in $\omega(k)$ in the $k \to 0$ limit, which in general required $x_{\text{as}}
\geq 2.2/k$. Here, by going to higher orders of the simplified hermitian theory,
we can determine quantities related to the error of the lowest order
approximation. The relatively simple eigenvectors (\ref{s0ex5}) with $g(x)
\equiv 1$ are not normalized which makes $Y_1(x)$ non vanishing. Normalization
of $\mathbf{s}_0(x)$ introduces a complication in the lowest order but
simplifies formulas for higher order corrections (due to $Y_1(x) \equiv 0$).
Using our present results, we can compare numerical values of the first and 2nd
order corrections in $Y(x)$ and $\mathbf{s}(x)$ at $x = x_{\text{as}}$ for the
normalized and non-normalized eigenvectors. These corrections tend to zero as
$x \to \infty$ and can be expected to fall below their values at $x =
x_{\text{as}}$. That is because the $\mathbf{R}$ matrix (\ref{Fex5}) tends to
a constant matrix as $x \to \infty$. In particular this implies
\begin{equation}
\label{limits}
\lim_{x \to \infty} \lvert Q(x) \rvert = \sqrt{1 + k^2 \pm \sqrt{1 + 4
\omega^2}}
\simeq \begin{cases} \sqrt{2} \\ k
\end{cases} \! , \quad \lim_{x \to \infty} \mathbf{s}_0(x) =
\{ 1, - 2 \omega \pm \sqrt{1 + 4 \omega^2} \} \simeq \{ 1, \pm 1 \} ,
\end{equation}
where the RHSs give the leading term in the $k \to 0$ limit
in which\cite{eias:BEC,asei:eigv} $\omega \simeq \frac{1}{2}k^2 \ln(1/k)$ also
tends to zero. The limiting value of $\lvert Q(x) \rvert$ for the lower sign is
small which is unfavorable for the PIA (as it makes $\epsilon_0$ large, see
Table~\ref{tab1}). This was the actual source of problems in the numerical
solving of the eigenvalue problem in question.
\begin{table}[h!]
\caption{\label{tab1}First and 2nd order corrections 
for non-normalized and normalized eigenvectors at $x=55$, $k=0.04$ and $\omega =
0.002604$. The first two lines represent $\mathbf{u}_{\text{ge}}$ and the last
two lines $\mathbf{u}_{\text{se}}$.}
\begin{ruledtabular}
\begin{tabular}{cccccc}
$\lvert Q \rvert$&$\epsilon_0/2$&$Y_1$&$Y_2$&$c_1^{\bot}$&$c_2^{\bot}$\\
\hline
1.41464\footnotemark[1] & -2.54639$\cdot10^{-8}$ & -8.4752$\cdot10^{-6}$ &
1.3783$\cdot10^{-7}$ & 
-1.70658$\cdot10^{-5}$ & -9.88846$\cdot10^{-7}$\\
1.41464\footnotemark[2] & -2.54639$\cdot10^{-8}$ & 0 & -2.55731$\cdot10^{-8}$ &
1.70658$\cdot10^{-5}$ & -9.88846$\cdot10^{-7}$\\
0.0427842\footnotemark[1] & 1.59832$\cdot10^{-2}$ & 2.83539$\cdot10^{-4}$ &
1.58104$\cdot10^{-2}$ & 
5.16137$\cdot10^{-7}$ & -3.15819$\cdot10^{-7}$\\
0.0427842\footnotemark[2] & 1.59832$\cdot10^{-2}$ & 0 & 1.59832$\cdot10^{-2}$ &
5.16137$\cdot10^{-7}$ & -3.15819$\cdot10^{-7}$
\end{tabular}
\end{ruledtabular}
\footnotetext[1]{Non-normalized eigenvector.}
\footnotetext[2]{Normalized eigenvector.}
\end{table}
The minimal value of $k$ which could be treated in our calculation was $k=0.04$.
The corresponding numerically found quantities were $\omega = 0.002604$ and
$C_{\text{se}}/C_{\text{ge}} = - 4.367$. To get $\omega$ practically independent
of $x_{\text{as}}$ required $x_{\text{as}} = 55$. For these values, the
contribution to $\mathbf{u}(x_{\text{as}})$ coming from the first term in
Eq.~(\ref{lincmb}) is nearly $26$ times larger than from the second term
($\lvert \mathbf{u}_{\text{se}}(x_{\text{as}}) \rvert/\lvert
\mathbf{u}_{\text{ge}}(x_{\text{as}}) \rvert \simeq \sqrt{\sqrt{2}/k} = 5.946$).
One can easily see that this fact also holds for the normalized eigenvector.
This means that the relative error in $\mathbf{u}(x_{\text{as}})$ is defined by
the relative error in $\mathbf{u}_{\text{se}}(x_{\text{as}})$. This in turn is
given by the corrections $Y_1(x_{\text{as}})$, $Y_2(x_{\text{as}})$ and the
multipliers $c_1^{\bot}(x_{\text{as}})$ and $c_2^{\bot}(x_{\text{as}})$, see
the last two lines in Table~\ref{tab1}. The dominating quantity is
$Y_2(x_{\text{as}})$. It depends very little on normalization. Therefore one
should not expect an improvement in the zero order approximation due to
normalization of $\mathbf{s}_0(x)$.

Note that in most cases, the second order corrections $Y_2(x_{\text{as}})$ are
very close to their values in the scalar case, $\epsilon_0(x_{\text{as}})/2$.

\section{\label{concl}Conclusions}

We present four generalizations of the Phase Integral Approximation to sets of
ODEs of the Schr\"odinger type (three for hermitian and one for the
non-hermitian sets).

The first is an extension of Fulling's hermitian theory \cite{Full} which
conserves the generalized current in each expansion order. In Ref.~8,
this theory was developed for positive definite matrices. In that case and for
real hermitian matrices, this theory conserves both the current associated with
the complex PIA and the Wronskian built from its real and imaginary part. Using
it for negative definite matrices makes the current vanish (and thus be
conserved). This can be of some interest only in such complex hermitian cases
with $Q^2 < 0$ in which in the simplified hermitian theory, the current given by
Eq.~(\ref{sigpia}) is not identically zero. This current is identically zero in
real hermitian cases where the $\mathbf{s}(x)$ vector is real, but also in
simple complex cases, see Sec.~\ref{exmp2} for examples.

The second generalization is the hermitian theory that conserves the Wronskian
built from two PIAs $\mathbf{u}^{\pm}(x)$. Our examples show that applicability
conditions for this theory are fulfilled through second order only if
$Q^2 > 0$. No such restrictions were found for negative definite matrices.
In that case and for real hermitian matrices, this theory conserves both the
Wronskian built from (real) $\mathbf{u}^{\pm}(x)$ and the current associated
with an approximate complex solution $\mathbf{u}^+(x) + i \, \mathbf{u}^-(x)$.

The third theory (``simplified hermitian'') conserves the current and the
Wronskian only in lowest order but contains no integrations characteristic of 
the first two. In the non-degenerate case, it contains no integrations in higher
order corrections and is thus fully algorithmic. Furthermore, these higher order
corrections were never found to be large far away from singularities. In such a
situation one can eliminate the small parameter $\lambda$ by putting $\lambda =
1$ and the idea of modification described in Sec.~\ref{auxfun} is applicable.
This theory is the simplest and in applications (like that described in
Sec.~\ref{realrhc}) may turn out to be the best. For $N=2$, using the program
in Mathematica described in Ref.~20,
one can determine corrections
of any (reasonable) order. No matter how complicated they are, their numerical
values can be determined. Writing a similar program for $N=3$ is not difficult.
For more than three equations, linear algebra becomes more complicated and
finding the eigenvalue $Q^2$ analytically may be impossible.

The non-hermitian theory for non-degenerate vector cases developed in
Sec.~\ref{nonht} is based on the same assumption as that in the simplified
hermitian theory ($(\mathbf{e}_1,\mathbf{s}_m) \equiv 0$). The example given in
Sec.\ref{nhc} seems to suggest that in typical situations ($Q^2 = x$), higher
order corrections are small at large distances from singularities ($x \to
\infty$). This theory is fully algorithmic and the idea of modification should
be applicable. Obviously, it can be used also for hermitian non-degenerate
matrices. In that case it gives the same results as the simplified hermitian
theory but without the necessity to introduce the basis in the orthogonal
complement of the eigenvector.

This paper only deals with the adiabatic part of the PIA for the vector case.
The connection problem \cite{FandF:book1,as:WKBcp,FandF:book2,FandF:book3}
requires a separate treatment. In the references just mentioned, this problem
for the scalar case was solved by tracing the unknown function $u(x)$ in the
complex $x$ plane. Note in this connection, that our non-hermitian theory
developed in Sec.~\ref{nonht} is the only vector theory of the PIA that
allows for complex $x$. Only within this theory can one try to solve the
connection problem for the vector case by tracing the unknown vector
$\mathbf{u}(x)$ in the complex $x$ plane.

\begin{acknowledgments}
I would like to dedicate this paper to the memory of my friend Per Olof
Fr\"oman, head of Department of Theoretical Physics at Uppsala University in
1964--1992, an expert in the theory and applications of the phase
integral approximation, who died suddenly last year. Our numerous discussions on
the phase integral approximation during my visits to Uppsala strongly influenced
my earlier results pertaining to the scalar case but also their generalizations
to the vector case given in this paper.

This work was initiated during our common research with E.~Infeld.
\cite{eias:BEC,asei:eigv} I would like to thank E.~Infeld for useful discussions
and help in preparing the manuscript. The referee was also helpful in pointing
out inconsistencies in Secs.~\ref{auxfun} and \ref{expls}, which have been
removed.
\end{acknowledgments}

\end{document}